\documentclass[aps,amsmath,amssymb,showpacs,showkeys]{revtex4-2}
\usepackage[dvips]{graphicx}
\usepackage{times}
\usepackage{braket}
\usepackage{xcolor}
\usepackage{orcidlink}
\usepackage{hyperref}
\hypersetup{
  colorlinks=true,
  urlcolor=magenta,
  linkcolor=red,
  citecolor=blue
}
\begin{document}
\title{Anisotropies of Diffusive Ultra-high Energy Cosmic Rays in $f(R)$ Gravity Theory}

\author{Swaraj Pratim Sarmah\orcidlink{0009-0003-2362-2080}}
\email[Email: ]{swarajpratimsarmah16@gmail.com}

\author{Umananda Dev Goswami\orcidlink{0000-0003-0012-7549}}
\email[Email: ]{umananda2@gmail.com}

\affiliation{Department of Physics, Dibrugarh University, Dibrugarh 786004, 
Assam, India}

%\date{}
\begin{abstract}
Understanding the anisotropy of ultra high-energy cosmic rays (UHECRs) is 
crucial for unraveling the origins and propagation mechanisms of these 
enigmatic particles. In this work, we studied the dipolar anisotropy of UHECRs 
in the diffusive regime by considering three cosmological models: the standard 
$\Lambda$CDM model, $f(R)$ gravity power-law model and the Starobinsky model.
This work aims to see the role of the $f(R)$ gravity theory in understanding 
the anisotropy of UHECRs without condoning the standard cosmology. We 
found that the amplitude of the dipolar anisotropy is sensitive to these 
cosmological models, with the $f(R)$ power-law model predicting the largest 
amplitude, while the $\Lambda$CDM model predicting the smallest amplitude at 
most of the energies in the range considered. The predicted amplitude of the 
Starobinsky model lies within the range of the $\Lambda$CDM one.
This work not only provides a way for exploration of UHECRs anisotropy within 
different cosmological contexts but also may pave the way for new avenues of 
research at the intersection of high-energy astrophysics.
\end{abstract}

%\pacs{}

\maketitle                                                                      

\section{Introduction}
Although it is thought that cosmic rays (CRs) are mostly of galactic origin 
below $10^{17}$eV and are most likely related to supernova remnants or 
pulsars \cite{blasi201321, berezhko661, hewitt2015}, their sources are still 
unknown. At ultra-high energies (UHEs) ($\geq 10^{18}$eV), CRs are 
most likely extragalactic in origin \cite{harari}. The main support for this 
theory comes 
from the fact that at energies of a few EeV ($1$ EeV $=10^{18}$eV), where the 
chemical composition is comparatively of low mass nuclei \cite{Auger2017}, 
the galactic magnetic field is not strong enough to cause CRs to diffuse within 
the galaxy and the arrival directions do not exhibit any noticeable 
correlation to the distribution of galactic matter \cite{Auger2013, TA2017}. Also, their extragalactic origin is supported by the 
detection of a dipolar distribution in the arrival directions of the CRs with 
energies above 8 EeV, which points away from the Galactic center direction 
\cite{Auger2017a}. The primary factors that help us to 
understand the properties of CRs are their energy spectrum, and anisotropies in 
their arrival directions which are measured at various angular scales. The 
changes in the slope of the spectrum can indicate alterations in 
propagation mechanisms or shifts in the source population, like the dominance 
of the extragalactic component above the ankle. These changes can also occur
at intermediate energies. Additionally, the impact of energy losses due to pair
production on extragalactic protons interacting with the cosmic microwave 
background (CMB) \cite{hillas677, blumenthal1596, berezinski_four_feat}, or diffusion effects \cite{lemoine_prd71}, can contribute to shaping the CR 
spectrum at energies in the EeV range. There are some key features in 
the UHECR spectrum: a slight increase in the intensity of the spectrum at the 
ankle around 5 EeV, then a softening of the spectrum at $\sim 13$~EeV
\cite{AugerPRL2020, AugerPRD2020}, followed by a significant drop starting at 
approximately 50 EeV \cite{Auger2021, icrc2021}.
 
Because of the incredibly high energies of CR's particles, they
surpass the energy that human-made accelerators can achieve. However, 
their arrival rate on Earth is meager, with only about one particle reaching 
the atmosphere per square kilometer per century with energies around or above 
$6 \times 10^{19}$ eV. Physicists are working hard to improve the accuracy and 
exposure to this tiny flux in order to unravel the mystery of these particles 
and their origins. Recent experimental advancements have been significant in 
this pursuit. One important discovery is that there 
is a decrease in the flux of CRs above $4 \times 10^{19}$ eV compared to
what was expected based on lower energy observations. This suppression has 
been confirmed by multiple studies \cite{AugerPRL2020, hires2008, Auger2008, 
TA2013L1}. The observed suppression in the flux above $4 \times 10^{19}$ eV 
could be due to the loss of energy during propagation over vast cosmological 
distances, a phenomenon predicted nearly fifty years ago 
\cite{KGreisen1966, GZ1966}. However, the current data is insufficient to 
determine if energy loss is the sole cause for this suppression. Researchers 
have also set upper limits on the presence of photons 
\cite{Auger2009, TA2013, auger_jcap05}, neutrinos 
\cite{Auger2012L4, Icecube2013, auger_jcap10}, and neutrons
\cite{Auger2012} among the UHECRs. Evidence from the Pierre Auger 
Observatory (PAO) suggests a shift from a lighter to a heavier composition
as the energy of CRs increases beyond $\sim 3 \times 10^{18}$ eV
\cite{Auger2010, Auger2013JCAP, auger_prd9012, auger_prd90}.

Physicists are still unsure about where UHECRs come from. However, 
they believe that by studying the anisotropies in their arrival directions, 
they may eventually figure it out \cite{mollerach201898, deligny_ptep2017, deligny_astrophys2019}. The challenge in this direction is that CRs, which are 
charged particles, are deflected by magnetic fields from their source 
directions in galactic or extragalactic spaces. Thus their arrival directions 
do not necessarily point to their sources. But as the rays gain more momentum, 
their deflection becomes less significant, especially towards the end of the 
highest energies observed. This gives hope that we can find the closest 
powerful sources outside our galaxy by looking for clusters of UHECRs pointing 
towards them. There are various possible types of extragalactic sources of 
UHECRs. The traditional viewpoint suggests that these sources could include 
powerful objects like gamma-ray bursts, tidal disruption events, active 
galactic nuclei, and galaxy mergers. It is important to consider how the 
brightness of these sources changes with distance (measured by redshift) 
because it affects both the weakening of CRs intensity at the highest 
energies and the creation of secondary particles through interactions with 
cosmic radiation backgrounds. These interactions can produce fragments from 
heavier CR particles or generate photons and neutrinos through photo-pion 
interactions. These studies have been conducted to investigate the 
importance of nearby sources in explaining the CR spectrum at extremely high 
energies. Additionally, it has been examined how these nearby sources may 
contribute to the observed patterns of anisotropy at intermediate angular 
scales. The PAO has observed the patterns of large-scale anisotropy in CR 
arrival directions. These patterns are naturally linked to the uneven 
distribution of galaxies in our nearby region, within a few hundred 
Mpc \cite{Auger2018}. Researchers have investigated the significance of nearby 
sources in explaining the spectrum of CRs at extremely high energies, as well 
as their potential role in accounting for anisotropies observed on 
intermediate angular scales \cite{blasi, taylor, matthews, guedes}.
The UHECRs detected by PAO on a large angular scale, analysed over 
three orders of magnitude in energy \cite{apj891}. For more than $4$ EeV, the 
dipolar amplitude increases with energy \cite{Auger2018}. Also, the researchers 
at the PAO have analyzed the arrival directions of more than $2600$ UHECRs 
above $32 \, \text{EeV}$, providing evidence for a deviation from isotropy at 
an intermediate angular scale with a $4\sigma$ confidence level for UHECR 
energies above $40\, \text{EeV}$ \cite{apj935}. The isotropy of UHECRs is 
disfavoured with $4\sigma$ confidence level in PAO observation with starburst 
galaxies, which is one of the significant results in UHECRs anisotropy 
\cite{apjl853}. The same kind of results can also be found in other giant 
CR experiments like Telescope Array 
\cite{TA_apjl898, TA_apj862, TA_arxiv14827, TA_auger_icrc}. In addition to 
studying the distribution of arrival directions of CRs with energy,
it also needs to understand how the composition of CRs changes with energy. 
Separating lighter and heavier components, which experience different levels 
of deflection, could also provide useful information and scientists are 
currently upgrading the Pierre Auger Observatory to help with these 
investigations.

The General Relativity (GR) developed by Albert Einstein in 1915 to 
explain gravitational interactions can be considered as the most beautiful, 
well-tested, and successful theory in this area. In 2015, the LIGO detectors 
\cite{ligo} detected the Gravitational Waves (GWs), which were predicted 
by Einstein almost a century earlier from his theory of GR. Additionally, in 
2019, the Event Horizon Telescope \cite{m87a, m87b, m87c, m87d, m87e, m87f}
released the first image of a supermassive black hole in the galaxy M87. 
These discoveries provide robust support for GR. However, GR has been 
suffering from some serious drawbacks also, such as the lack of a complete 
quantum theory and its inability to explain the current accelerated expansion 
of the Universe \cite{reiss, perlmutter, spergel, astier} as well as the 
missing mass \cite{Oort, Zwicky1, Zwicky2, Garrett, nashiba1} in galaxies' rotational dynamics. To address these issues, Modified Theories of Gravity (MTGs) have been developed along with the concept of dark energy \cite{sami, udg_prd}. 
One widely used MTG is the $f(R)$ theory of gravity \cite{Sotiriou}, where the 
Ricci scalar $R$ in the Einstein-Hilbert action is replaced by a function 
$f(R)$. In recent times various models of $f(R)$ gravity, i.e.\ the functional 
forms of $f(R)$ have been proposed to explain these cosmic phenomena. Some of 
extensively used and viable models of $f(R)$ gravity are the Starobinsky 
model \cite{starobinsky, staro}, Hu-Sawicki model \cite{husawicki}, power-law 
model \cite{powerlaw, udg_ijmpd}, and Tsujikawa model \cite{tsujikawa}.

Various research groups have used various methods to study CR 
anisotropy and propagation till now \cite{mollerach2022, harari, molerach, 
harari2021, merksch, m.ahlers, mollerech2022b, Abeysekara, grapes3, 
globus2019, G_sigl1999}. Considering the significant contributions of MTGs in 
understanding cosmological \cite{psarmah, gogoi_model} and astrophysical 
\cite{jbora, nashiba2,ronit1, nashiba3} issues in recent
times, it is prudent to explore the application of MTGs in the field of CRs to 
address the current challenges in this domain. With this motivation, in this 
work, we are interested in studying CR anisotropies in the energy range between 
$10^{17}$ eV and $10^{20}$ eV in the realm of MTGs for the very first time. In 
our previous work, we studied flux and propagation properties of UHECRs in 
the $f(R)$ theory of gravity \cite{swaraj}. In that study, we here
considered two very well-known $f(R)$ gravity models, viz., the power-law model 
\cite{powerlaw} and the Starobinsky model \cite{staro}. In this study also we 
use these two models of $f(R)$ gravity along with the standard $\Lambda$CDM
model for comparison. The chief aim of our study is to see the effect of $f(R)$ 
gravity theory on the anisotropic behaviour of UHECRs in comparison to the 
standard cosmological perspective, rather than justifying the correctness of 
theory in terms of observational data of this interesting behaviour of CRs.   

The rest of the parts of this paper are arranged as follows. 
Since in this study, we consider the $f(R)$ gravity theory as the basic 
cosmological theory required to study the propagation of UHECRs in galactic
and extragalactic spaces, in Section \ref{secII} we discuss briefly the 
$f(R)$ gravity models used in this study and also present the required 
cosmological equations obtained for those models. Section \ref{secIII} is 
divided into two parts. Subsection \ref{subsecIIIA} is dedicated for the 
theoretical formalisms of the propagation of UHECRs in turbulent magnetic 
fields (TMF) and hence their consequent dipolar anisotropy. The numerical 
calculations of the anisotropies of UHECR protons along with a few nuclei 
have been analysed and discussed in Subsection \ref{subsecIIIB}. Finally, we 
summarize our results and give final remarks in Section \ref{secIV}.

\section{$f(R)$ gravity models and cosmological equations}
\label{secII}
In this study, we use two most prevalent and viable $f(R)$ gravity 
models: the power-law model and the Starobinsky model. It should be mentioned 
at this point that two $f(R)$ models are used to gain confidence in the 
obtained results from the comparison and also to understand their performance 
in this sector of CR physics. Here we briefly introduce these two models and
also write the expressions of the Hubble parameter $H(z)$ for them. These 
expressions of $H(z)$ will be used to calculate different parameters for
both models. The details about these two models and derivations of the 
expressions of $H(z)$ for them can be obtained in 
Refs.~\cite{staro, powerlaw, udg_ijmpd, swaraj}.

The functional form of the power-law model is 
given by \cite{powerlaw,udg_ijmpd}
\begin{equation}\label{powerlaw}
f(R)=\sigma\, R^n,
\end{equation}
where ${\sigma}$ and $n$ are two free model parameters. The 
best-fitted value of the model parameter $n$ is found as 1.4 \cite{powerlaw}. 
The parameter ${\sigma}$ relies on $H_0$, $n$ and $\Omega_{m0}$ as 
given by
\begin{equation}\label{lambda}
{\sigma} = -\,\frac{3H_0^2\, \Omega_{m0}}{(n-2)R_0^n}\,.
\end{equation}
The expression of the Hubble parameter $H(z)$ for this model can be written 
as \cite{powerlaw}
\begin{equation}\label{powerlawhubble}
H(z) = \left[-\,\frac{2nR_0}{3 (3-n)^2\, \Omega_{m0}} \Bigl\{(n-3)\Omega_{m0}(1+z)^{\frac{3}{n}} + 2 (n-2)\,\Omega_{r0} (1+z)^{\frac{n+3}{n}} \Bigl\}\right]^\frac{1}{2}\!\!,
\end{equation}
where $R_0$ is the present value of the Ricci scalar and is given by 
\cite{powerlaw}
\begin{equation}\label{R0}
R_0 = -\, \frac{3 (3-n)^2 H_0^2\, \Omega_{m0}}{2n\left[(n-3)\Omega_{m0} + 2 (n-2) \Omega_{r0}\right]}\,.
\end{equation} 

Our study aims to elucidate the correlation between the redshift of a 
remote celestial entity and its age. This objective can be achieved through 
an exploration of the relationship between redshift and the evolution of 
cosmological time, which is given by
\begin{equation}\label{dtdz}
\bigg | \frac{dt}{dz} \bigg |=\frac{1}{(1+z)\,H(z)}.
\end{equation}
%The expression for $H(z)$ will be used to calculate different parameters for 
%both the power-law model and the Starobinsky model.

Using Eq.~\eqref{powerlawhubble}, we can write Eq.~\eqref{dtdz} for 
the power-law model as
\begin{equation}\label{dtdz1}
\bigg | \frac{dt}{dz} \bigg | = (1+z)^{-1}  \left[-\,\frac{2nR_0}{3 (3-n)^2 \Omega_{m0}} \Bigl\{(n-3)\Omega_{m0}(1+z)^{\frac{3}{n}} + 2 (n-2)\Omega_{r0} (1+z)^{\frac{n+3}{n}} \Bigl\} \right]^{-\,\frac{1}{2}}\!\!\!\!\!\!\!.
\end{equation}

Again for the Starobinsky model, we consider the functional form as \cite{staro}
\begin{equation}\label{starobinsky}
f(R) = \alpha R + \beta R^2.
\end{equation}
Here $\alpha$ and $\beta$ are two model parameters. The best-fitted values of 
these model parameters are found to be 1.07 and 0.00086, respectively 
\cite{swaraj}. The expression for the Hubble parameter for the Starobinsky 
model can be obtained as \cite{swaraj}
\begin{equation}\label{hubble_staro}
H(z) = H_0 \left[ \frac{3\, \Omega_{m0} (1+z)^3 + 6\, \Omega_{r0} (1+z)^4 + 
\left(\alpha R + \beta R^2\right)\!H_0^{-2}}{6(\alpha + 2 \beta R)\Bigl\{ 1-\frac{9\,\beta H_0^2\, \Omega_{m0} (1+z)^3}{\alpha(\alpha+2\beta R)} \Bigl\}^2 }\right]^{\frac{1}{2}}\!\!\!.
\end{equation}
And thus using the above expression of $H(z)$, we can write the 
Eq.\ \eqref{dtdz} explicitly for the Starobinsky model as
\begin{equation}\label{dtdz_staro}
\bigg | \frac{dt}{dz} \bigg | = \big[(1+z) H_0\big]^{-1} \left[ \frac{3\, \Omega_{m0} (1+z)^3 + 6\, \Omega_{r0} (1+z)^4 + \frac{\alpha R + \beta R^2}{H_0^2}}{6(\alpha + 2 \beta R)\Bigl\{ 1-\frac{9\beta H_0^2 \Omega_{m0} (1+z)^3}{\alpha(\alpha+2\beta R)} \Bigl\}^2 }\right]^{-\frac{1}{2}}\!\!\!\!\!\!,
\end{equation}

The expressions of $|dt/dz|$ for both the cosmological models can be used to 
calculate the density enhancement factor, modification factor, CR flux etc., 
which are discussed in detail in Ref.\ \cite{swaraj}. We use the Hubble 
constant $H_0 \approx 67.4$ kms$^{-1}$ Mpc$^{-1}$ \cite{planck2018}, matter 
density parameter $\Omega_{m0} \approx 0.315$ \cite{planck2018} and radiation 
density parameter $\Omega_{r0} \approx 5.373 \times 10^{-5}$ \cite{nakamura} 
in our numerical calculations. 

\section{Anisotropy of Ultra-High Energy Cosmic Rays}\label{secIII}
In this section, we delve into the exploration of UHECRs' anisotropy 
in the light of two models of $f(R)$ theory of gravity, the power-law model 
and the Starobinsky model as mentioned in the previous section, considering
the propagation of UHECRs in the presence of TMFs. This section is bifurcated 
into two subsections for a systematic detailed analysis. The theoretical 
formalism for examining the anisotropy of UHECRs in the presence of a TMF in 
the $f(R)$ theory of gravity is developed in the first subsection. The second 
subsection is devoted to the numerical analysis of the obtained theoretical 
formulation and to the related discussions.

\subsection{Theoretical Formalism}
\label{subsecIIIA} 
It is believed that the evolution of primordial seeds $\sim$ 1 nG impacted by 
the process of structure building may generate the TMFs in the Universe at 
present \cite{hu_apj, urmilla}. Magnetic fields with some strength 
associated with matter density are often augmented in dense places, such as 
superclusters. Extragalactic magnetic fields can also be produced by galactic 
outflows, in which galactic magnetic fields are transported into the 
intra-cluster medium by winds. Although magnetic fields with $\mu$G strengths 
have been recorded in cluster cores, they are projected to be smaller at 
supercluster scales, and values ranging from 1 nG to 100 nG have been 
studied \cite{feretti, Valle, Vazza}, with the assumed coherence length 
$l_c$ of the order of 0.1 -- 1 Mpc \cite{sigl}. We will consider the 
individual CR sources in the local supercluster, which is a group of galaxies 
that includes our own Milky Way galaxy. The local supercluster is located 
within 100 Mpc of Earth. For simplicity, we will assume that a uniform, 
isotropic TMF is present within the diffusion region, which is the region of 
space where CRs can propagate. The field will be characterized by a root mean 
square strength $B = \sqrt{\langle B^2(x)\rangle}$. An effective Larmor radius 
for charged particles can be defined as
\begin{equation}\label{larmor}
r_L = \frac{E}{ZeB} \simeq 1.1\, \frac{E/\text{EeV}}{ZB/\text{nG}}\;\text{Mpc}.
\end{equation}

The Larmour radius is equal to the coherence length at the critical 
energy $E_c$, which distinguishes between resonant diffusion at low energies 
and non-resonant diffusion at high energies, i.e. $l_c = r_L (E_c)$. The critical energy of the particles 
is a significant factor in the study of the diffusion of charged particles in 
magnetic fields and it is given by
\begin{equation}\label{cri_energy}
E_c = ZeBl_c \simeq 0.9 Z\, \frac{B}{\text{nG}}\, \frac{l_c}{\text{Mpc}}\;\text{EeV}.
\end{equation}
The critical energy $E_c$ distinguishes between two regimes of CRs diffusion, 
resonant diffusion at low energies ($<E_c$) and non-resonant diffusion at high 
energies ($>E_c$). In the resonant diffusion regime, CRs are deflected by 
magnetic field fluctuations with scales comparable to the Larmour radius. In 
the non-resonant diffusion regime, deflections are smaller and can only occur 
over distances greater than $l_c$. The diffusion coefficient D as a function of energy is given by
\cite{harari}
\begin{equation}\label{diff_coeff}
D(E) \simeq \frac{c\,l_c}{3}\left[4 \left(\frac{E}{E_c} \right)^2 + a_I \left(\frac{E}{E_c} \right) + a_L \left(\frac{E}{E_c} \right)^{2-m}   \right],
\end{equation}
where $c$ is the speed of light. In the case of the Kolmogorov spectrum 
$m=5/3$, $a_L \approx 0.23 $ and $a_I \approx 0.9$, and for that of Kraichnan 
spectrum $m = 3/2$, $a_L \approx 0.42 $ and $a_I \approx 0.65$. The density 
${\rho}$ of relativistic particles propagating from a source that lies at 
${\bf x}_s$ in an expanding Universe obeys the equation during the diffusion 
phase as \cite{berezinkyGre}
\begin{equation}\label{diff_eqn}
\frac{\partial {\rho}}{\partial t} + 3 H(t)\,{\rho} - b(E,t)\,\frac{\partial {\rho}}{\partial E}-{\rho}\, \frac{\partial {\rho}}{\partial E}-\frac{D(E,t)}{a^2(t)}\,\nabla^2 {\rho} = \frac{Q_s(E,t)}{a^3(t)}\,\delta^3({\bf x}-{\bf{x}_s}),
\end{equation}
where $H(t)= \dot{a}(t)/a(t)$ is the Hubble parameter, $a(t)$ is the scale 
factor, ${\bf x}$ describes the comoving coordinates, $Q_s(E)$ is a source 
function which denotes the number of particles that are emitted with energy $E$ per unit time. The energy losses of the emitted particles are described by
\begin{equation}
\frac{dE}{dt} = -\, b(E,t),\;\; b(E,t) = H(t)E + b_{int}(E).
\end{equation}
This comprises energy redshift due to cosmic expansion and energy losses due 
to interactions with radiation backgrounds, including pair production and 
photo-pion generation as a result of interactions with the CMB (for details see 
\cite{harari}). For the case of zero energy losses, the solution of Eq.\ 
\eqref{diff_eqn} is given by
\begin{equation} \label{solution0}
{\rho}(r_s, t, E) = \frac{Q(E)\,\exp \left[-r_s^2/4 Dt \right]}{(4 \pi Dt)^{3/2}},
\end{equation}
where $r_s$ is the source distance.
Now, integrating this Eq.\ \eqref{solution0} over time, we obtain the solution 
as
\begin{equation}
\boldsymbol{\rho}(r_s, E) = \frac{Q(E)}{4 \pi r_s D(E)}.
\end{equation}

The general solution of Eq.\ \eqref{diff_eqn} including the energy losses have 
been taken into account was given by \cite{berezinkyGre} as
\begin{equation}\label{density}
\boldsymbol{\rho}(E,r_s)= \int_{0}^{z_{i}} dz\, \bigg | \frac{dt}{dz} \bigg |\, Q(E_g,z)\, \frac{\exp \left[-r_s^2/4 \lambda^2\right]}{(4\pi \lambda^2)^{3/2}}\, \frac{dE_g}{dE},
\end{equation}
where $z_i$ is the initial redshift and $E_g$ is the generation energy having 
energy $E$ at the redshift $z=0$. The parameter $dt/dz$ is the 
relation between cosmological time evolution and the redshift as we have 
already discussed in Section \ref{secII}. $\lambda$ is the Syrovatsky 
variable \cite{syrovatsy_1959, mollerach2013} and it is given as
\begin{equation}\label{syrovatsky}
\lambda^2(E,z)=\int_{0}^{z}dz\, \bigg | \frac{dt}{dz} \bigg |\,(1+z)^2 D(E_g,z).
\end{equation}
The variable $\lambda(E, z)$ represents the conventional distance covered by 
CRs originating at redshift $z$ with energy $E_g$, from their point of 
generation to the current moment when their energy has decreased to $E$. We 
are interested in the expression that quantifies the rate at which the energy 
of particles at their source degrades about their energy at $z = 0$, 
denoted as $dE_g/dE$ and is given by \cite{berezinkyGre, berezinski_four_feat}
\begin{equation}\label{degde}
\frac{dE_g}{dE}= (1+z)\, \exp \left[ \int_{0}^{z} dz\, \bigg | \frac{dt}{dz} \bigg | \left(\frac{\partial\, b_{int}}{\partial E} \right) \right].
\end{equation} 
The detailed derivation of $dE_g/dE$ was nicely performed by Berezinsky 
et al.~in Appendix B of Ref.~\cite{berezinski_four_feat}. In the following we 
will implement the power-law and Starobinsky models' results from 
Ref.\ \cite{swaraj} to obtain the CR's protons density enhancement factor, and 
subsequently the CR's protons flux and finally the CR anisotropies as 
predicted by these two $f(R)$ gravity models. We calculate the dipolar 
anisotropy using Ref.\ \cite{Supanitsky} as
\begin{equation} \label{aniso}
\Delta = 3 ~ \frac{\eta}{\xi},
\end{equation}
which is identical to the expression in Ref.\ \cite{harari} (according to 
\cite{Supanitsky}). Here, $\eta$ and $\xi$ are the modification factor and
enhancement factor respectively. The analytical expressions for $\eta$ and 
$\xi$ are taken from Refs.\ \cite{molerach, swaraj} and accordingly we will 
proceed to calculate it for the $\Lambda$CDM model, power-law model and the 
Starobinsky model. It needs to be mentioned that at sufficiently high 
energies, CR enters the quasi-rectilinear regime, where only angular 
diffusion occurs but there is no role of the spatial diffusion. In our work, 
the considered anisotropy Eq.~\eqref{aniso} which depends on the modification 
factor and enhancement factor is valid for a wide range of energies 
(see Ref. \cite{Supanitsky}). The reason for choosing this model is that it 
is compatible with different cosmological models. Although the 
quasi-rectilinear regime is certainly important and relevant at sufficiently 
high energies, it falls outside the scope of our current study. Further, the 
$\Lambda$CDM model is used here as a standard model for comparison with the 
$f(R)$ gravity models' results. 

\begin{figure}[!h]
\centerline{
\includegraphics[scale=0.328]{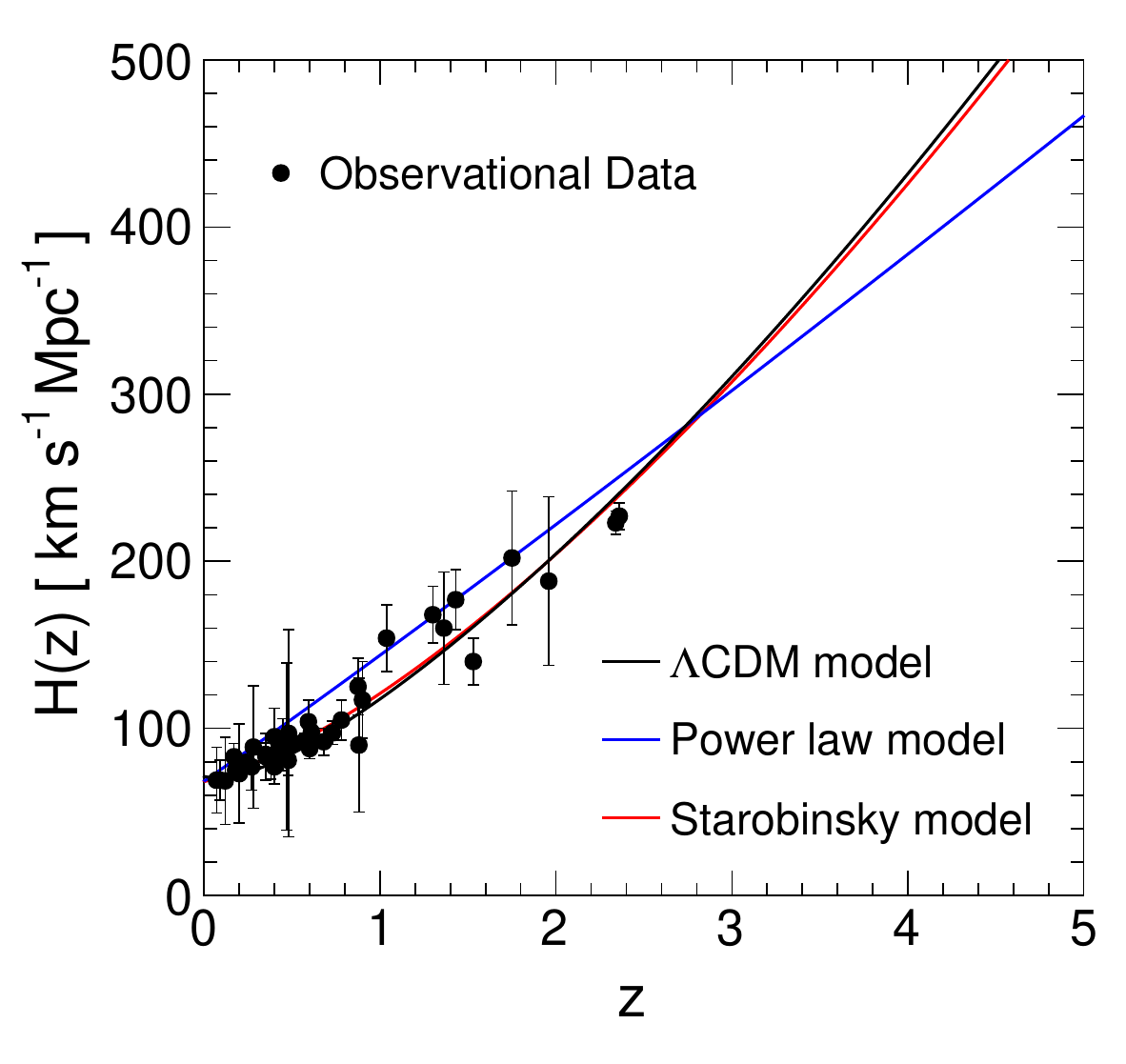} \hspace{1cm}
\includegraphics[scale=0.32]{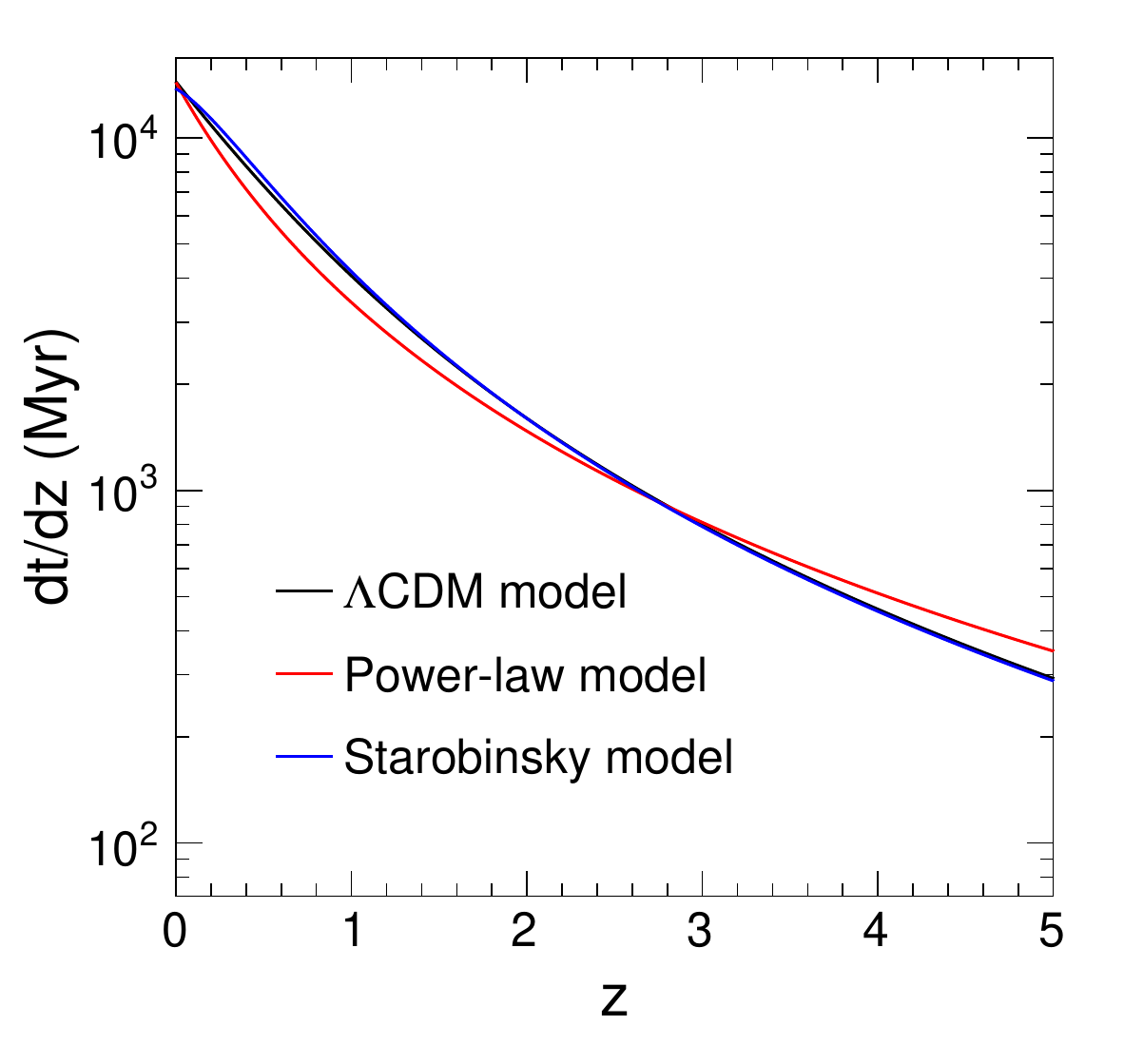}}
\vspace{-0.2cm}
\caption{Hubble parameter $H(z)$ with the observational Hubble dataset (left panel) and $dt/dz$ (right panel) as a function of the redshift $z$ for the $\Lambda$CDM, power-law and the Starobinsky model.}
\label{fig1_new}
\end{figure}

It would be pertinent to mention at this point that the dipolar 
anisotropy of UHECRs is influenced by the distribution of their sources and 
their propagation behaviours in the intergalactic medium. The propagation 
behaviours are determined by the cosmological parameters, which differ in 
different cosmological models. Each of these models has a different set of 
cosmological parameters, which leads to different propagation effects and 
hence different predictions for the dipolar anisotropy. To present this 
point clearly, we show the plots of $H(z)$ and $dt/dz$ with respect to $z$ 
for all three considered models in Fig.~\ref{fig1_new}. One can see from these 
plots that both $f(R)$ gravity models along with the $\Lambda$CDM model 
satisfy the available $H(z)$ observational data. This demonstrates that these 
$f(R)$ gravity models are consistent with observations, despite the 
differences in their predictions of $H(z)$ values at different $z$. It is 
seen that there are overlappings of all three models at $ z\sim$ 
$0.029$ and $2.8$ within the considered range of $z$. Whereas, except around 
these overlapping points there is a noticeable difference in the predictions 
of the power-law model from the other two models. Power-law model predicts 
lower values of ${dt/dz}$ (higher values of ${H(z)}$) than that for 
the other two models from ${z \sim}$ $0.029$ to ${z \sim} $ $2.8$. 
For ${z}\gtrsim$ $2.8$ the power-law model predicts higher values of 
${dt/dz}$.  
  
\subsection{Numerical Results and Discussions}\label{subsecIIIB}
\begin{figure}[b]
\centerline{
\includegraphics[scale=0.38]{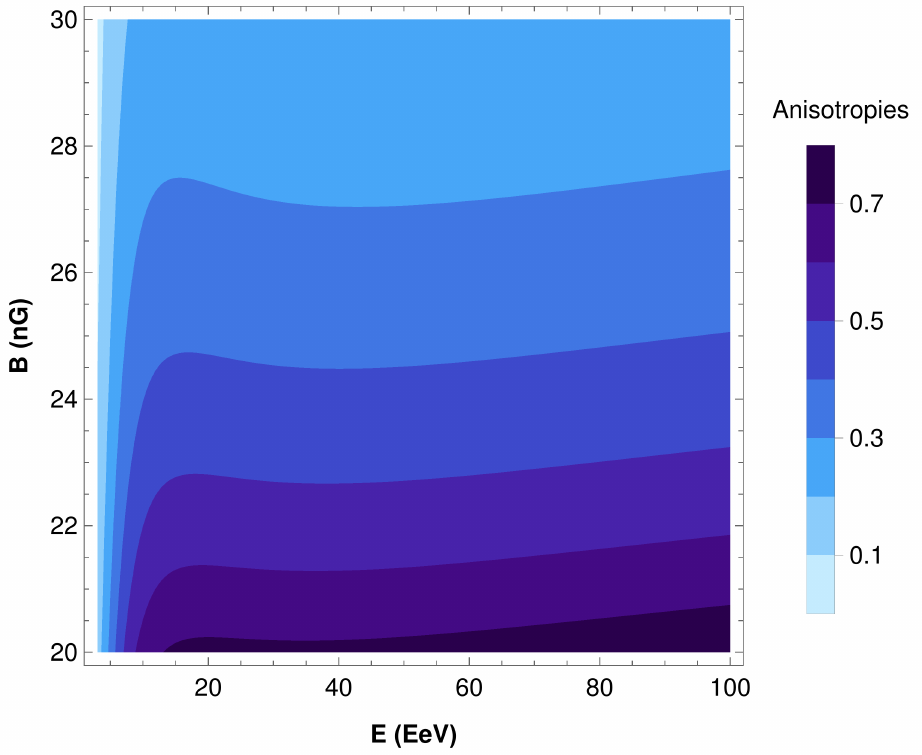}
\includegraphics[scale=0.38]{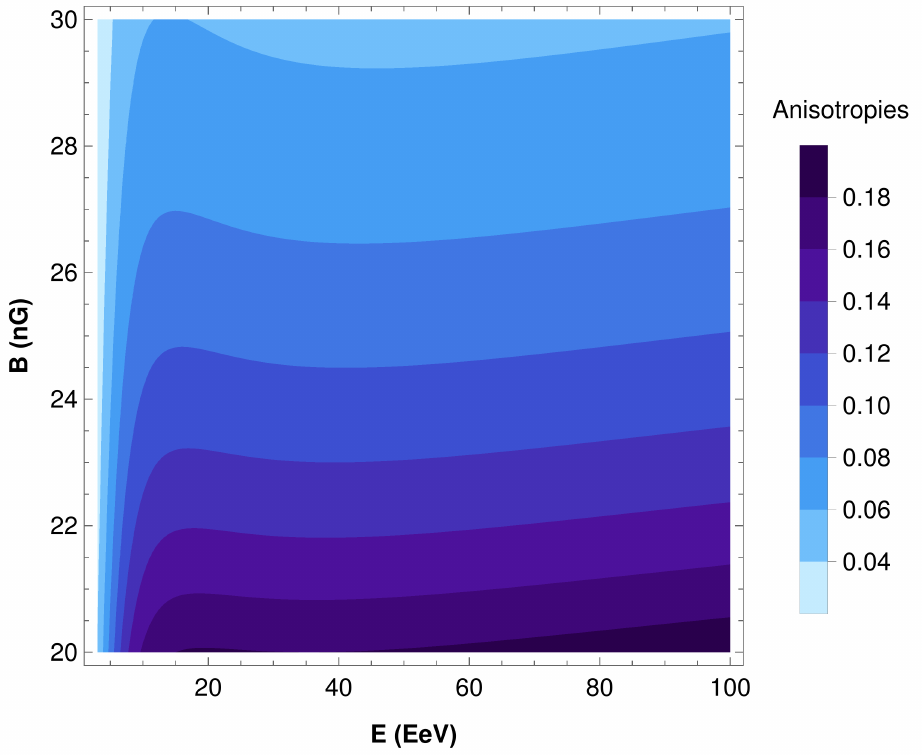}
\includegraphics[scale=0.38]{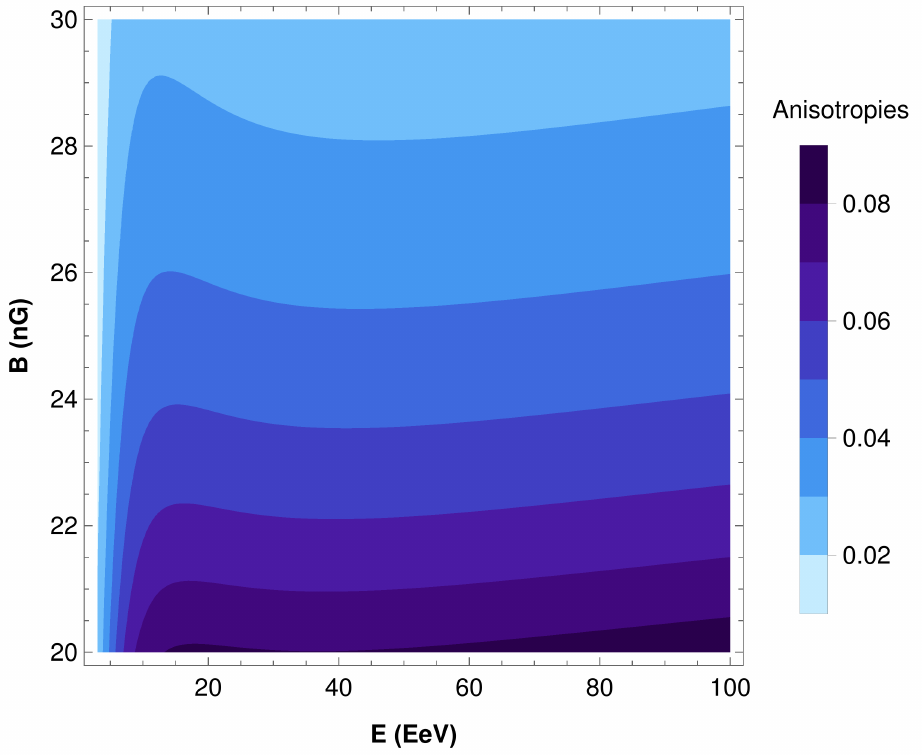}}
\centerline{
\includegraphics[scale=0.38]{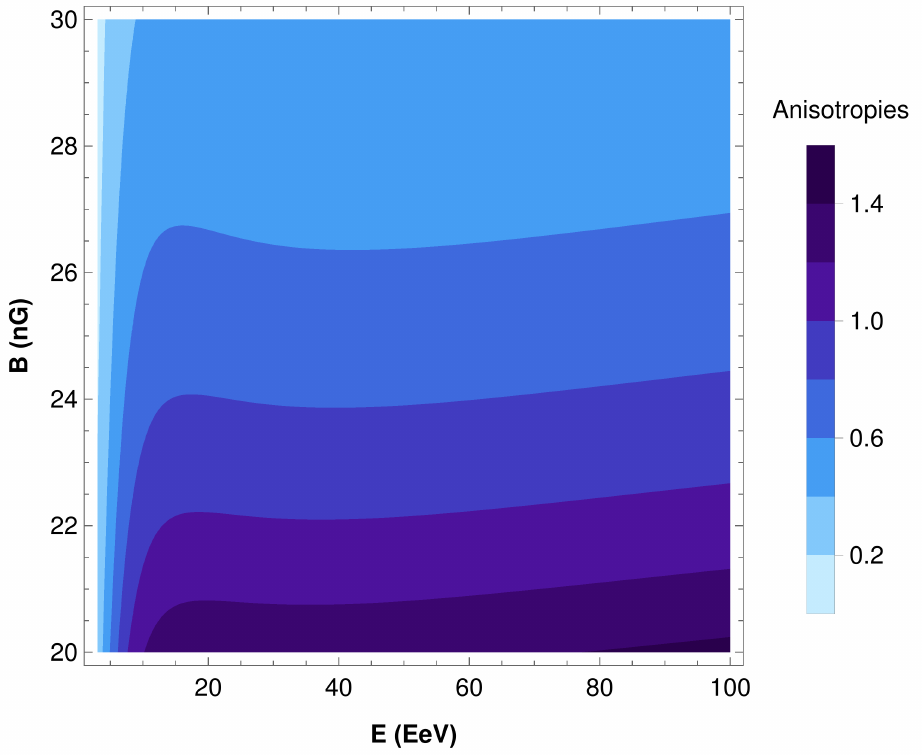}
\includegraphics[scale=0.38]{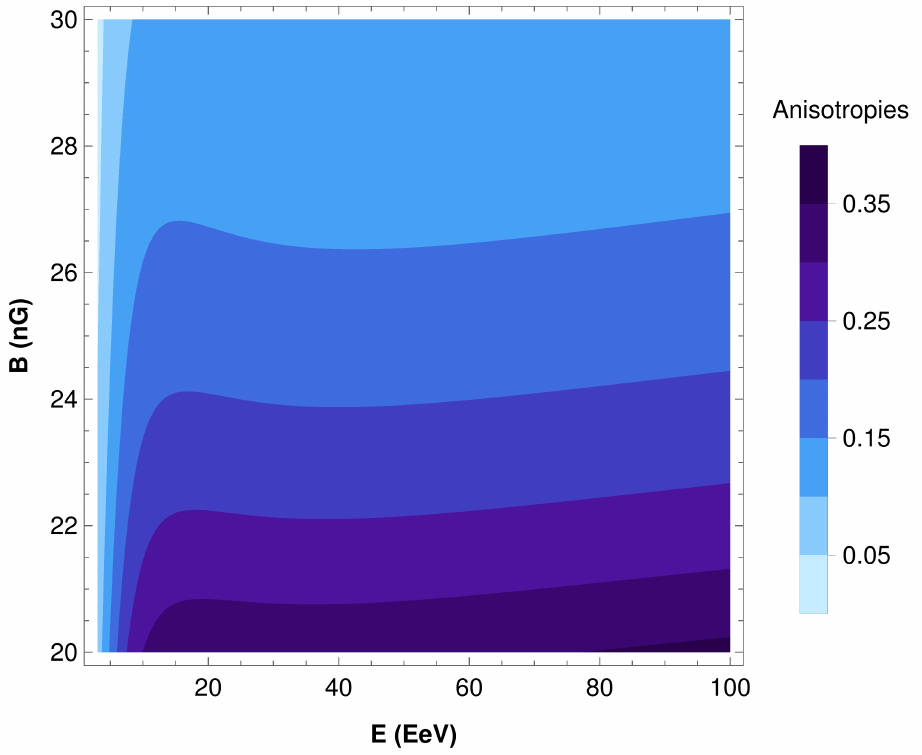}
\includegraphics[scale=0.38]{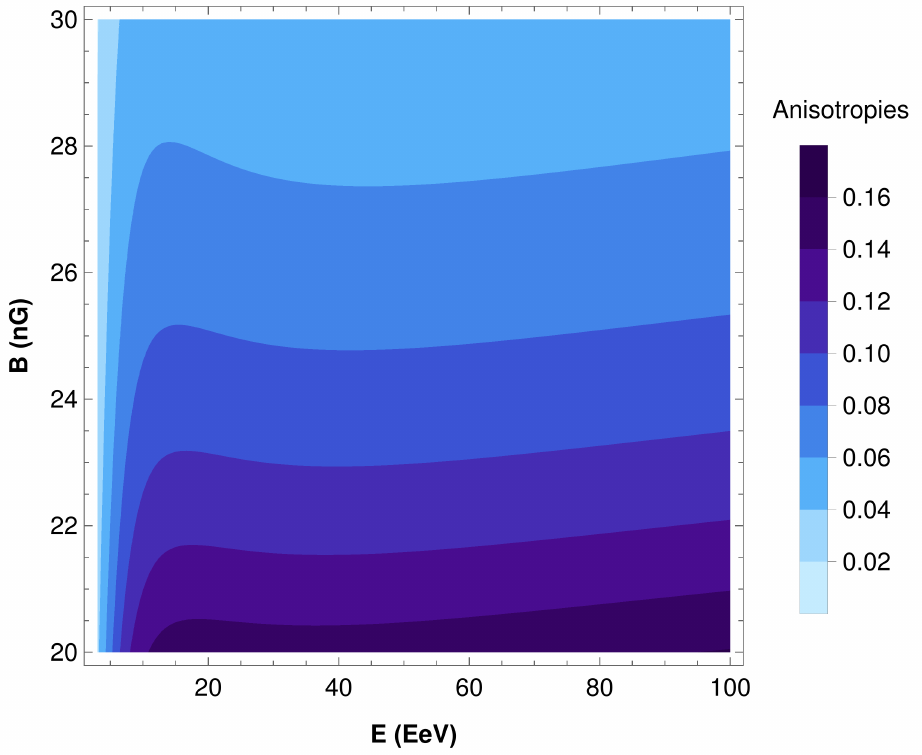}} 
\centerline{
\includegraphics[scale=0.38]{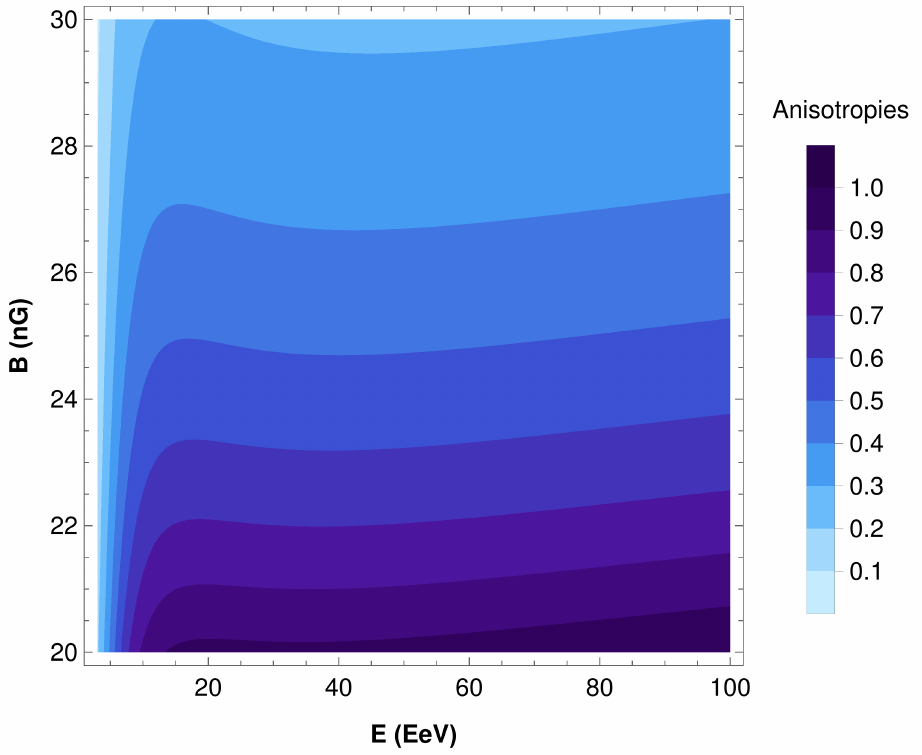}   
\includegraphics[scale=0.38]{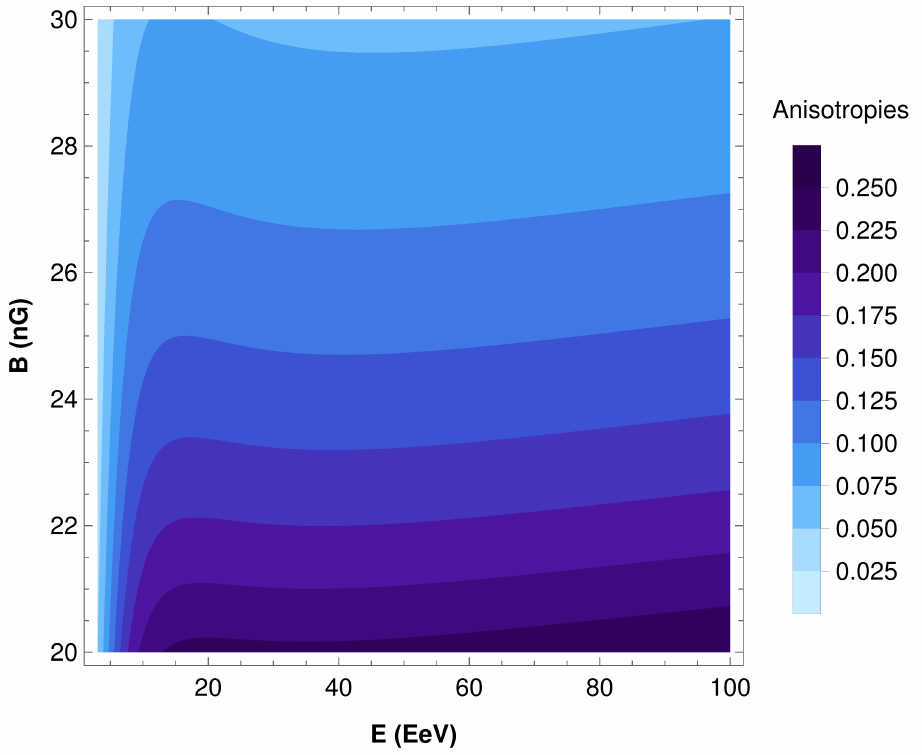}
\includegraphics[scale=0.38]{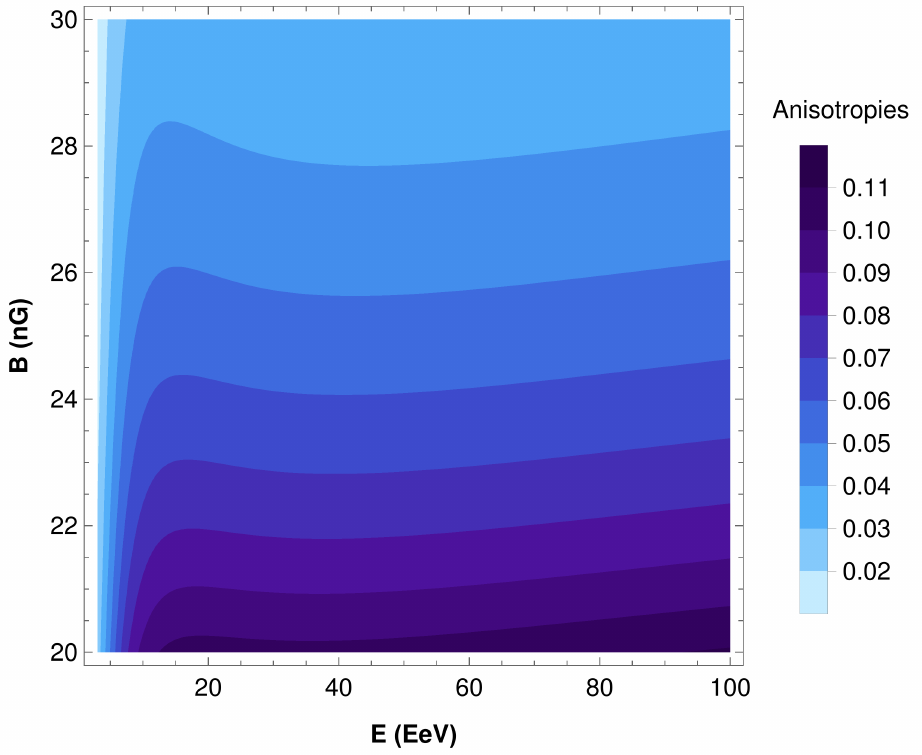}} 
\caption{Anisotropies of UHECRs for the $\Lambda$CDM model (top panels), 
$f(R)$ power-law model (middle panels) and the Starobinsky model 
(bottom panels) at selected source distances of $30$, $60$, and $90$ Mpc 
(from left to right) for each of the models. Here for simplicity 
composition of CRs is considered as pure proton.}
\label{fig1}
\end{figure}
Using Eq.\ \eqref{aniso} we calculate the anisotropies of UHECRs for the 
considered $f(R)$ gravity models: the power-law model and the Starobinsky 
model, at selected source distances of 30, 60, and 90 Mpc considering the 
composition of UHECRs as pure proton for simplicity. The same calculations 
are also done for the standard $\Lambda$CDM model for the comparative 
analysis. The results of these calculations are shown as contour plots in 
Fig.\ \ref{fig1}. For the calculation purpose, unless specified 
otherwise, we take the coherence length  $l_c=0.25$ Mpc. This 
considered coherence length lies within its range of values used in the 
literature as mentioned earlier. It is to be noted that the enhancement 
factor $\xi$ is directly proportional to $l_c$. So increasing $l_c$ leads to 
an increase in the enhancement factor, hence decreasing the anisotropy. 
The contours show the levels of anisotropy at different 
energy and magnetic field values. The color scale in each plot represents 
the level of anisotropy, from high at the top to low at the bottom. It is seen 
that for a given magnetic field, the CR anisotropy increases as the energy 
increases, which is an established fact. This is because the deflection of 
CRs by the galactic magnetic field decrease with increasing energy. At low 
energies, the deflections are large enough to randomize the arrival directions 
of the CRs, making them appear isotropic. On the other hand, at high energies, 
the deflections are small enough that the CRs can still retain some of their 
directional information. This leads to an increase in the anisotropy at high 
energies.
\begin{figure}[h!]
\centerline{\hspace{-0.28cm}
\includegraphics[scale=0.61]{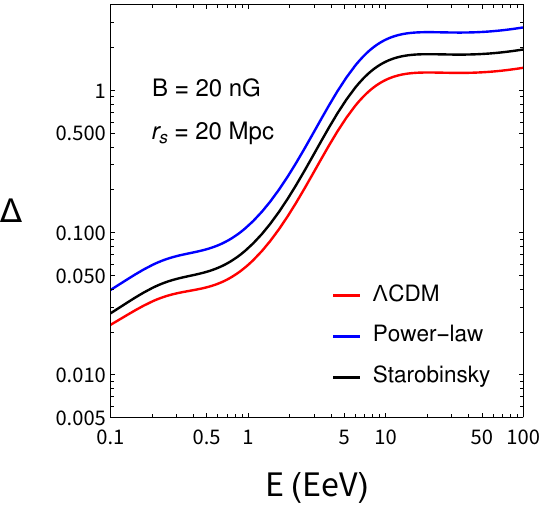} \hspace{1.2cm}
\includegraphics[scale=0.61]{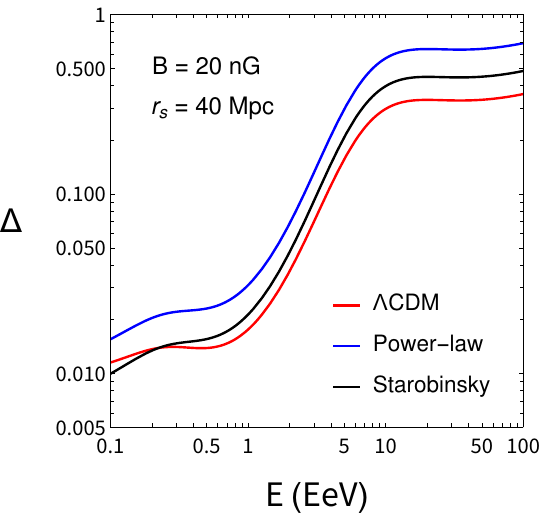}}\vspace{0.3cm}
\centerline{
\includegraphics[scale=0.58]{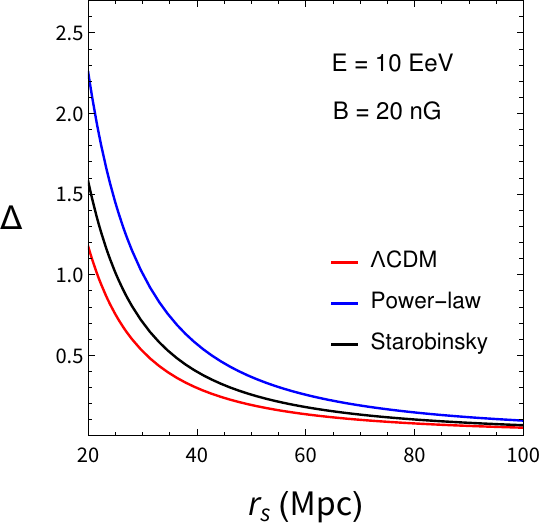} \hspace{1.5cm}
\includegraphics[scale=0.58]{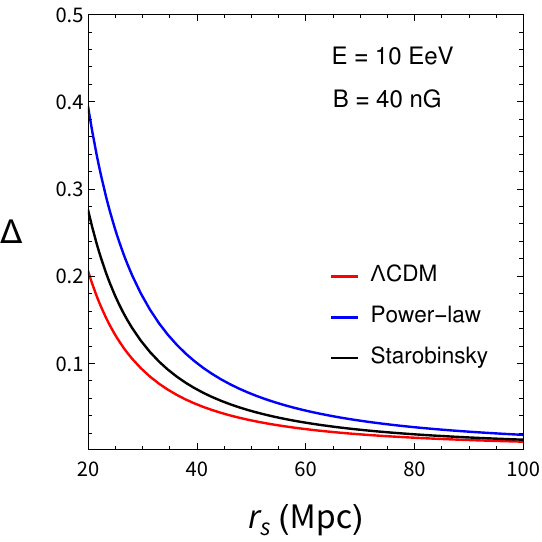}}\vspace{0.3cm}
\centerline{
\includegraphics[scale=0.58]{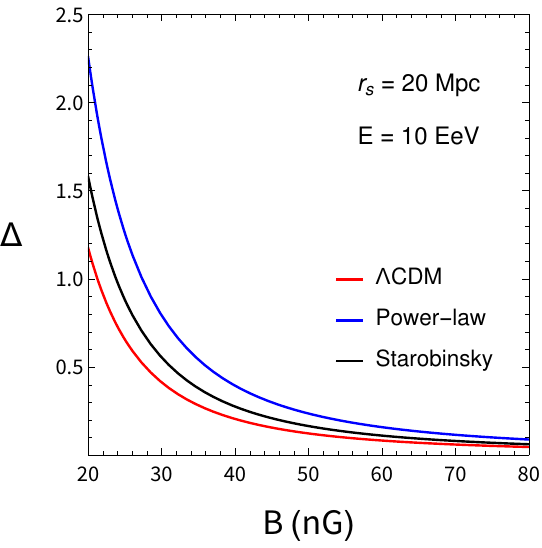} \hspace{1.5cm}
\includegraphics[scale=0.58]{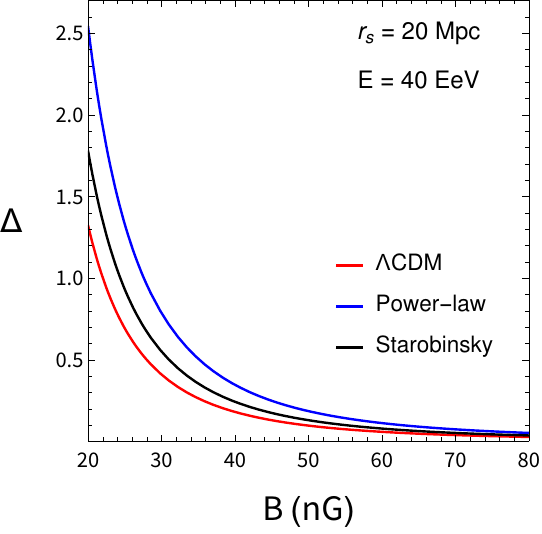}}
\caption{Comparison of UHECRs' anisotropy as a function of energy (top panels),
source distance (middle panels) and  magnetic field strength (bottom
panels) for different cosmological models.}
\label{fig2}
\end{figure}

\begin{figure}[h!]
\centerline{
\includegraphics[scale=0.6]{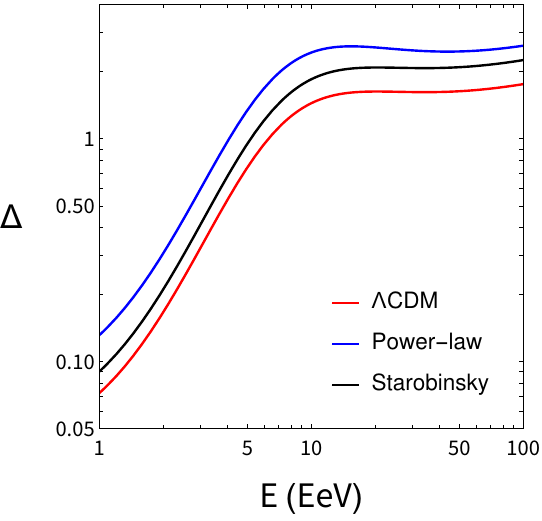} \hspace{1.5cm}
\includegraphics[scale=0.374]{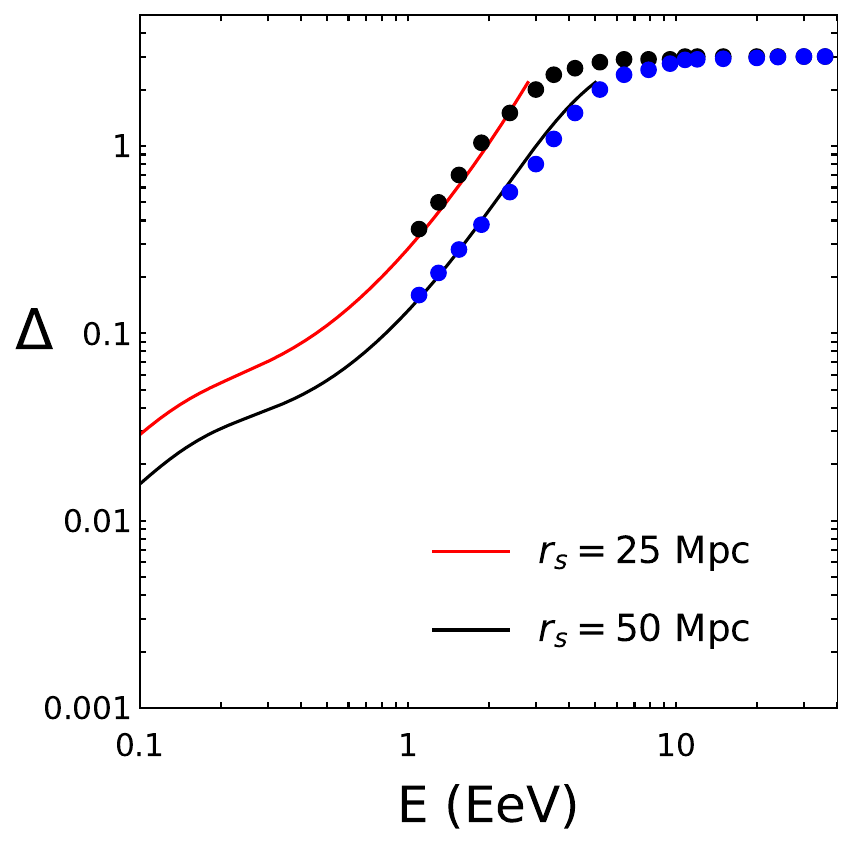}}
\caption{Left: Variation of UHECRs' anisotropy with respect to the 
energy for the $\Lambda$CDM model, $f(R)$ power-law model and the Starobinsky 
model. Right: The anisotropy obtained for the $\Lambda$CDM model of this 
present work (solid lines) along with the results obtained from the 
stochastic differential equation as shown in Ref.~\cite{harari}}
\label{fig4_new}
\end{figure}

\begin{figure}[h!]
\centerline{
\includegraphics[scale=0.38]{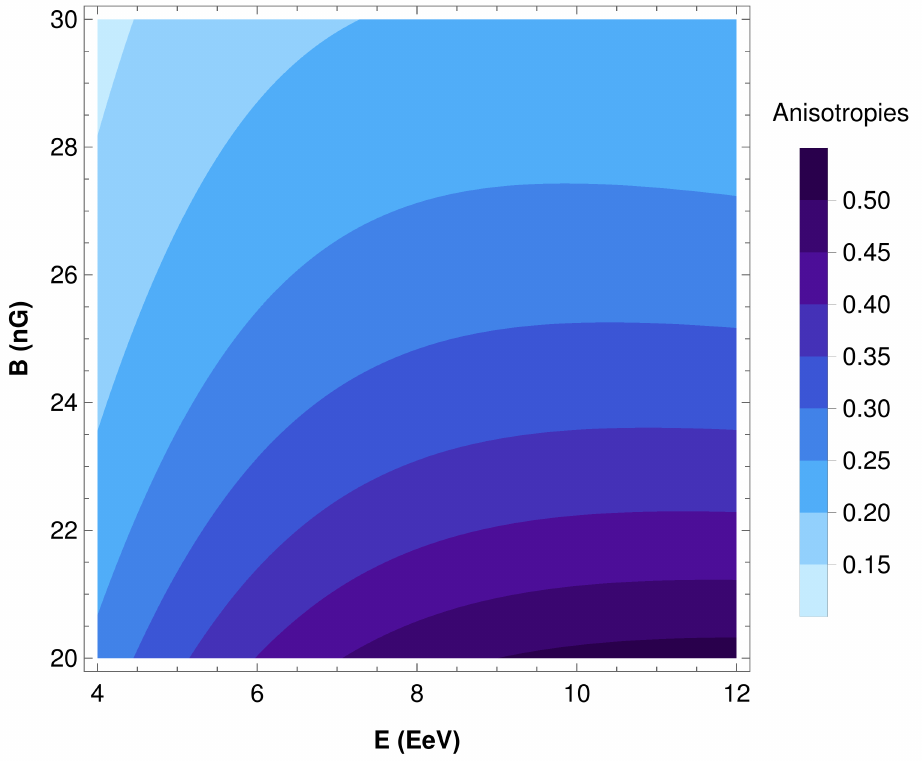}
\includegraphics[scale=0.38]{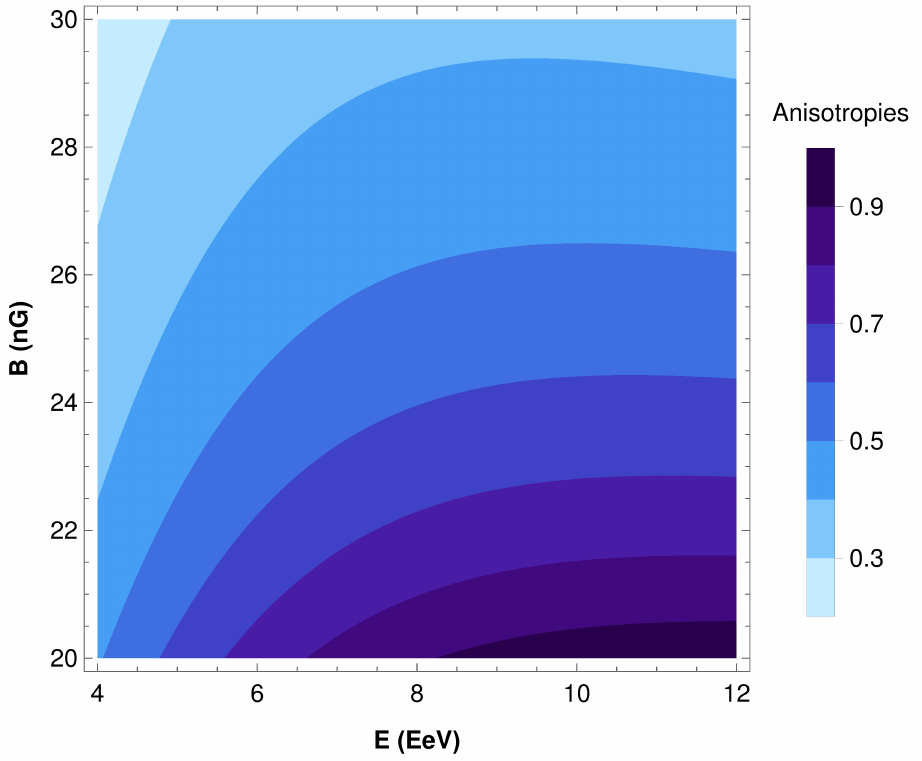}
\includegraphics[scale=0.38]{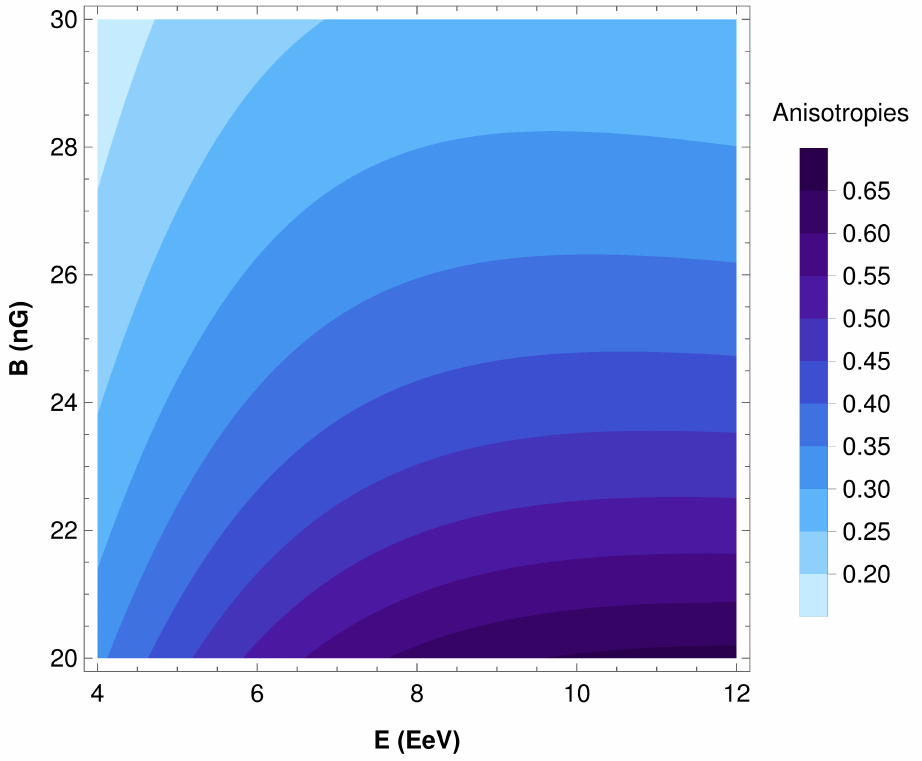}}
\centerline{
\includegraphics[scale=0.38]{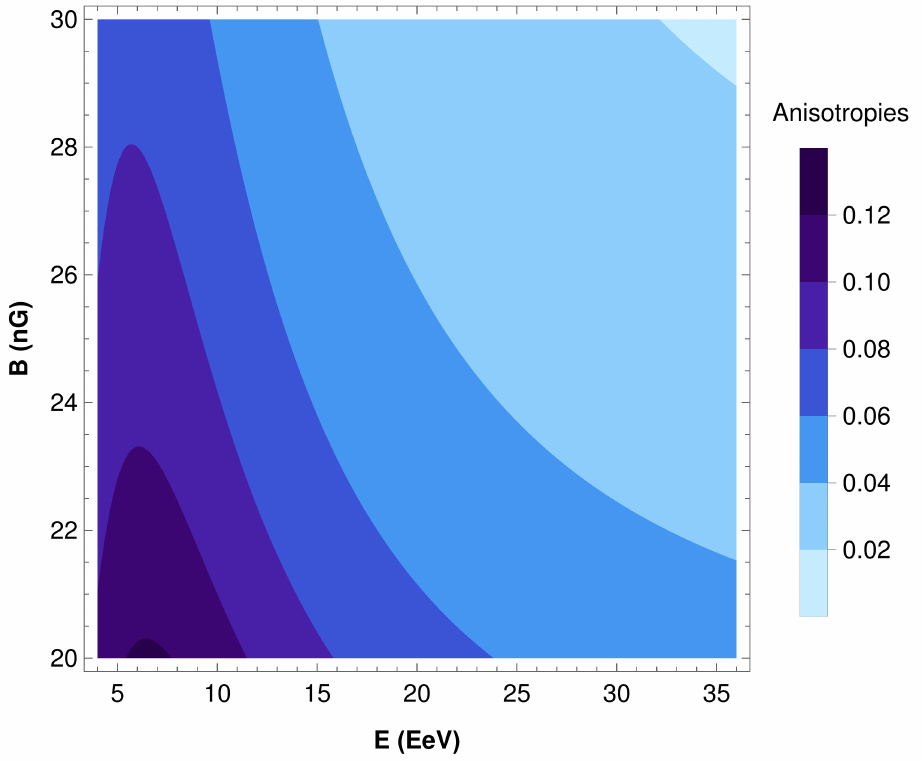}
\includegraphics[scale=0.38]{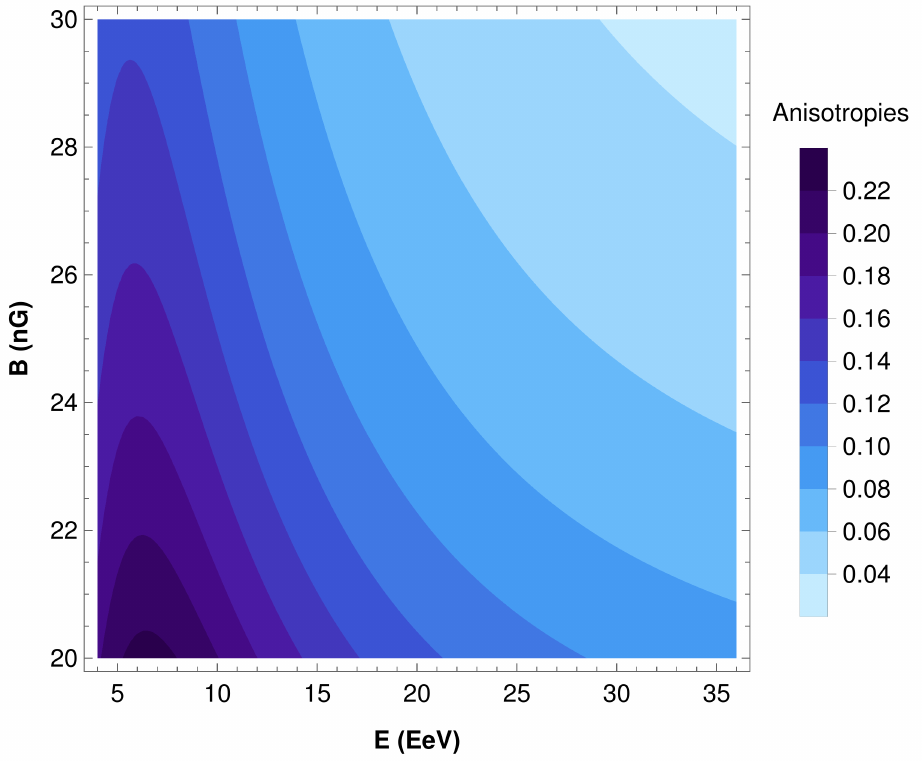}
\includegraphics[scale=0.38]{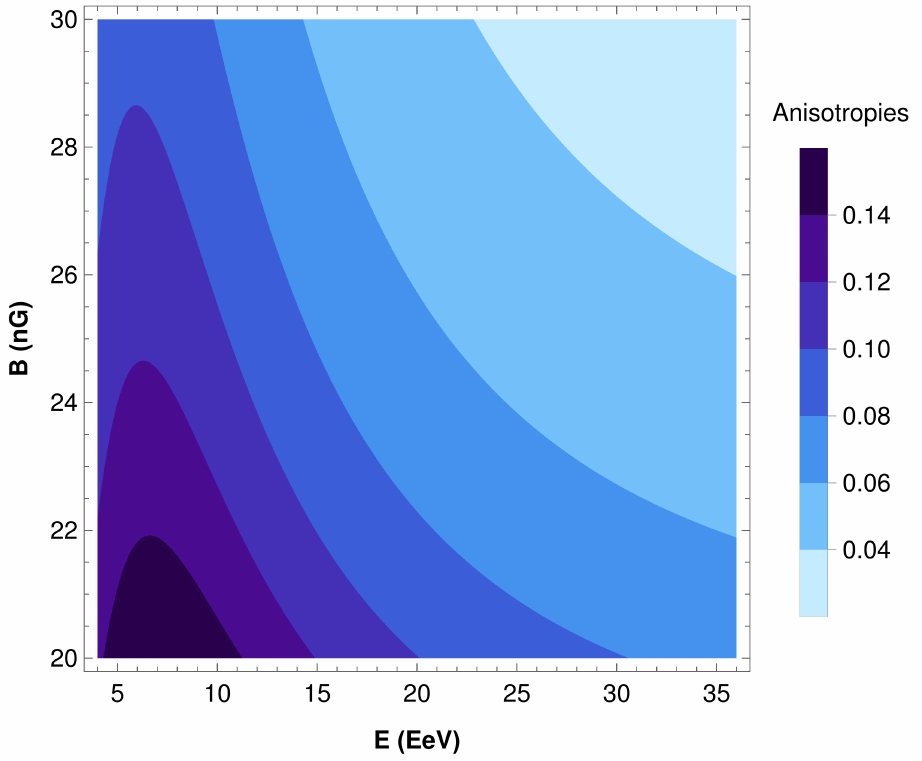}} 
\centerline{
\includegraphics[scale=0.38]{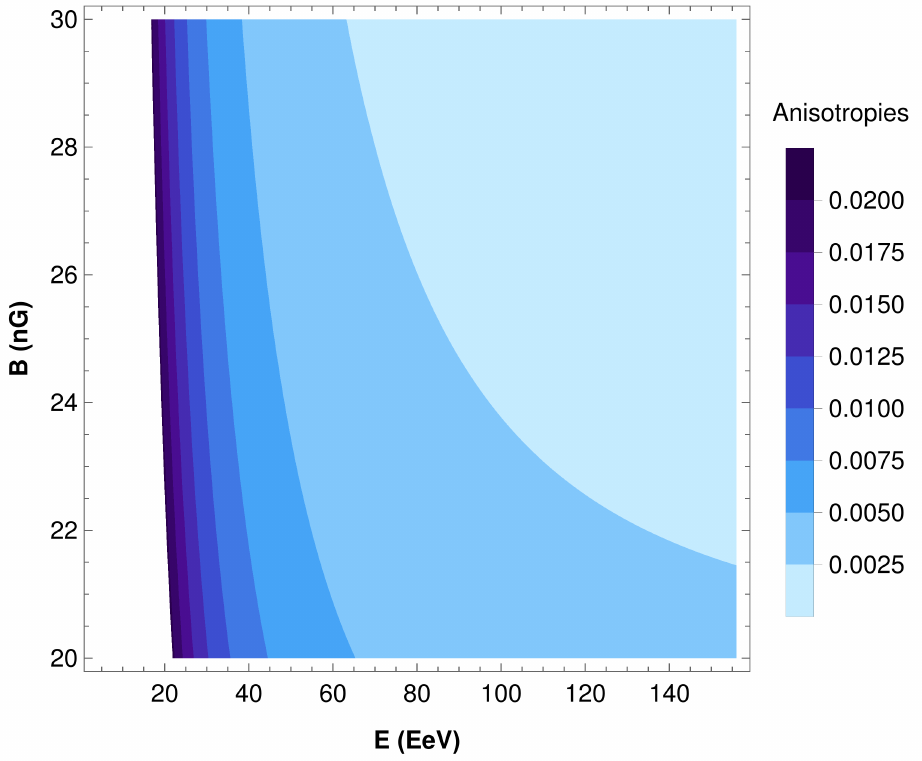}   
\includegraphics[scale=0.38]{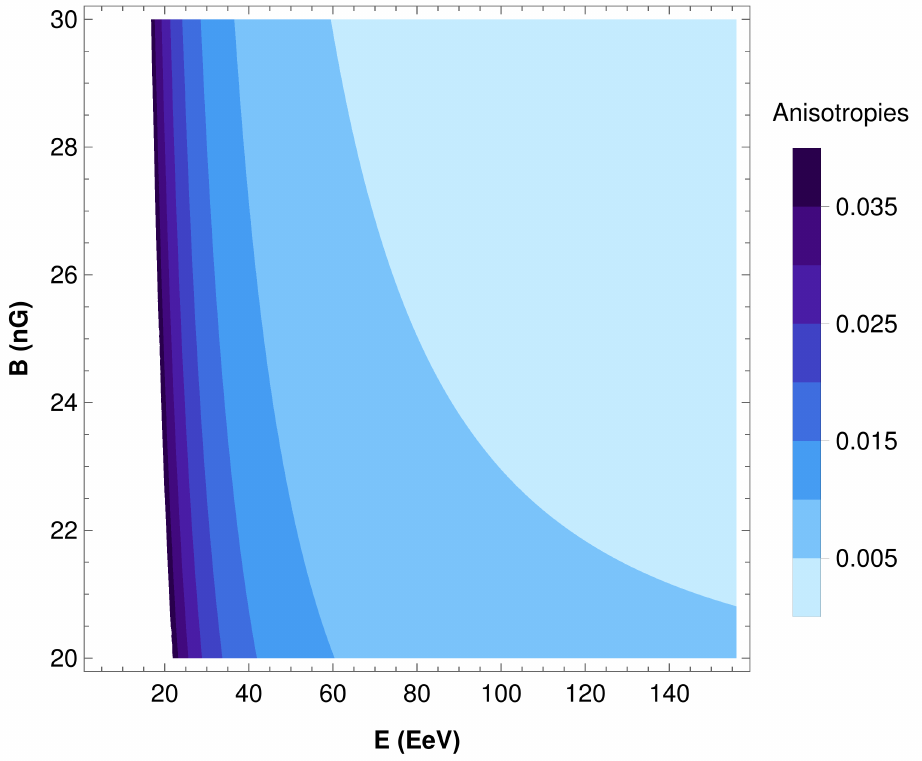}
\includegraphics[scale=0.38]{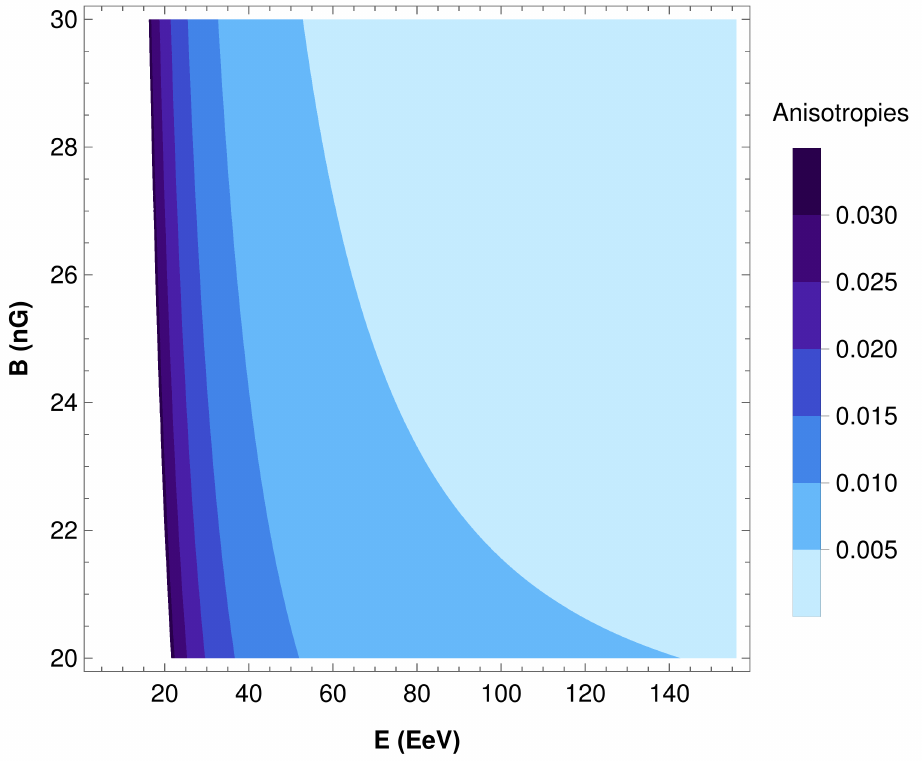}} 
\caption{UHECRs' anisotropy for pure helium (top panels), carbon 
(middle panels) and iron (bottom panels) nuclei for the $\Lambda$CDM model, 
$f(R)$ power-law model and the Starobinsky model (vertical panels from left 
to right) respectively at 30 Mpc source distance.}
\label{fig3}
\end{figure}  

The results of the plots show that the CRs' anisotropy is strongly 
dependent on both energy and magnetic field. At high energies and less 
strength of the magnetic field, the anisotropy is very high, with a level of 
about $0.02$ to $1.4$ depending on the cosmological model as 
well as the source distance considered. This means that CRs of high energies 
in a low magnetic field region are much more likely to come from a particular 
direction than CRs of low energies in a region of a high magnetic field. In 
contrast, at high energies and higher strength of the
magnetic field, the anisotropy is quite low, with a level of about 
$0.03$ to $0.48$ depending on the considered model and the 
source distance. This means that CRs even at high energies with 
the effect of stronger magnetic fields are much more isotropic, or evenly 
distributed in all directions. These results of the plots can be explained 
from the process of diffusion of CRs in the galactic or extragalactic
magnetic field. As mentioned already, it is obvious that at low energies, CRs 
are more easily scattered by the magnetic field, which reduces their 
anisotropy. But other factors such as source distance and strength of magnetic 
field also have effects on their anisotropy simultaneously. At high energies, 
CRs are not easily scattered by the magnetic field, which allows them to 
travel in straighter lines and maintain their anisotropy. However, a much 
stronger magnetic field can scatter even UHECRs sufficiently to have isotropic 
distribution as indicated by the plots. We also draw the contour plots for 
different source distances in Fig.~\ref{fig1}. It is seen that as the distance from the sources 
increases, the anisotropy of CRs decreases. As a whole, the $\Lambda$CDM model 
shows the least anisotropies among others, while the $f(R)$ power-law model 
shows much higher values of anisotropy. The Starobinsky model shows the 
anisotropy values within the $\Lambda$CDM model and the power-law model. Thus 
the $f(R)$ gravity models give the anisotropy values that are in a wider range 
than that of the $\Lambda$CDM model. 

To specify some of the points clearly as discussed above with some 
additional features, we draw the UHECRs' anisotropy with respect to energy, 
source distance, and strength of the magnetic field in Fig.~\ref{fig2}. Each
plot of the figure is obtained by keeping any two fixed at their specific 
values while taking one as a variable from energy, source distance, and 
magnetic field strength. The different curves in the plots of the 
figure represent different cosmological models. It is to be noted that 
the $\Lambda$CDM is the model that is most commonly used in CR studies. As 
mentioned earlier the $f(R)$ power-law model and the Starobinsky model are 
the models that we propose to study the behaviour of anisotropy of UHECRs. 
From the figure, we see that at a given source distance and the 
magnetic field value (see top plots), the power-law model depicts the highest 
values of anisotropy for the whole range of energies considered (within the UHE 
scale). While the $\Lambda$CDM model depicts the lowest anisotropy values 
for the whole energy range, especially at a small source distance. However, at 
a long source distance, the Starobinsky model predicts the lowest anisotropy in
the lower energy range. For example at $r_s = 40$ Mpc, the Starobinsky model 
gives the lowest anisotropy values for energies $E< 0.3$ EeV. This range
of energy may become wider by shifting towards the higher energy side for 
more strong magnetic fields. Only 
above this energy range, the $\Lambda$CDM model depicts the lowest values in 
this case. Moreover, in the lower energy range, the $\Lambda$CDM model and the 
Starobinsky model depict the same anisotropy value at a particular energy 
depending on the source distance and magnetic field strength. As an example, 
for $r_s=40$ Mpc and $B=20$ nG, the same anisotropy value is predicted by 
these two models at $\sim$ $0.3$ EeV. The middle panels show the variation of 
anisotropy of UHECRs with the source distance for the two sets of fixed 
values of energy and magnetic field strength. Here the anisotropy decreases 
with increasing the source distance as expected and the power-law model 
predicts maximum anisotropy values throughout the considered range of source 
distances. Lastly, the bottom panels show the behaviour of anisotropy of 
UHECRs with the magnetic field strength for the two sets of fixed 
values of energy and source distance. As anticipated it is seen that 
anisotropy decreases with an increase in magnetic field strength. It is also 
seen from here that anisotropy increases with an increase in energy as observed 
earlier. In this case also the power-law model gives the highest anisotropies 
followed by the Starobinsky model and $\Lambda$CDM model. Nevertheless, it is 
to be noted that the difference in prediction of anisotropy by these 
cosmological models decreases with an increase in both source distance and the 
magnetic field strenth. This is due to the reason that as the anisotropy 
decreases with increasing source distance and the magnetic field strength, the 
difference of prediction of anisotropy by cosmological models also decreases.

Still, to lucidly visualise the effect of the cosmological models on
the UHECRs' anisotropy, we plot the anisotropy with respect to energy from
$1$ EeV to $100$ EeV considering a small source distance $r_s=10$ Mpc and a 
very weak magnetic field $B=1$ nG on the left panel of Fig.~\ref{fig4_new} 
for all three cosmological models considered. From this plot, one can see that 
even in this scenario also there is a considerable effect of the cosmological 
models on the anisotropy of UHECRs over the whole range of energy considered. 
This is because each model has a different set of model parameters due to 
which it provides a different effective Hubble parameter value from the others 
as mentioned earlier (see Fig.~\ref{fig1_new}) and hence provides a different 
anisotropy value than the rest.

For the comparison of our findings with the results of available
literature, we consider the results of Ref.~\cite{harari} in which only 
$\Lambda$CDM model was taken into consideration. In this context, it needs 
to be mentioned that we consider the energy spectral index $\gamma=2.7$ and 
coherence length $l_c = 0.25$ Mpc with a range of magnetic field values in our 
work. Whereas in Ref.~\cite{harari} $\gamma=2$ and $l_c=1$ Mpc have been used. 
Thus for verifying our results with that of the Ref.~\cite{harari}, we calculate
the dipolar anisotropy values of UHECRs within the corresponding energy range 
with $\gamma=2$, $B=1$ nG and $l_c=1$ Mpc for the $\Lambda$CDM model only. The 
results are as shown in the right panel of Fig.~\ref{fig4_new} in comparison 
to that of Ref.~\cite{harari} for two source distances of $25$ Mpc and 
$50$ Mpc. The dots shown in the plot are obtained from the integration of the 
stochastic differential equation as shown in Ref.~\cite{harari}, while the 
solid lines represent the anisotropy from our results and show a very good 
agreement.
        
Further, we draw the contour plots of UHECRs' anisotropy considering the nuclei as the UHECRs composition in Fig.\ \ref{fig3}. 
It is to be noted however that the formation of secondary nucleons during 
photo-disintegration processes affects nuclear masses, making it challenging 
to account for energy losses in the case of nuclei \cite{harari}. For such a 
nuclear composition, one has to replace $E$ by $E/Z$, where $Z$ is 
the atomic number. The anisotropy levels of 
these nuclei are quite different from the pure proton case. Here, we again 
consider the $\Lambda$CDM model, power-law model, and the Starobinsky model 
taking the source at a distance $r_s=30$ Mpc. We consider He, C and Fe 
nuclei with a $E^{-2.7}$ spectral distribution by taking the energy range from 
$4$ EeV upto $6Z$ EeV \cite{harar_2015_prd}. 
For instance, in the He nuclei case, the energy range is from $4$ EeV to $12$ 
EeV. We see that the anisotropy increases with increasing energy and decreasing 
magnetic field strength. In the case of the carbon nuclei, the 
variations of anisotropy are more quicker with energy and magnetic field, 
while for the iron nuclei, the lower value of anisotropy covers most of the 
range and also they have a rapid variation of anisotropy than the carbon 
nuclei. These behaviours of carbon and helium nuclei also have some dependency 
on the cosmological model considered. The small values of anisotropy and their 
behaviours in the case of considered nuclei may be due to the attribution of 
the process of photo-disintegration as mentioned above, for which heavier 
nuclei are likely to be more prone. Again, if we take a look at the 
cosmological models considered here, the power-law model depicts the highest 
value of anisotropy, while the $\Lambda$CDM model shows the lowest value. 
The Starobinsky model shows the results between the $\Lambda$CDM model 
and the power-law model. The higher anisotropy is obtained for 
lighter nuclei as compared to the heavier nuclei \cite{harar_2015_prd}.

\begin{figure}[h!]
\centerline{
\includegraphics[scale=0.38]{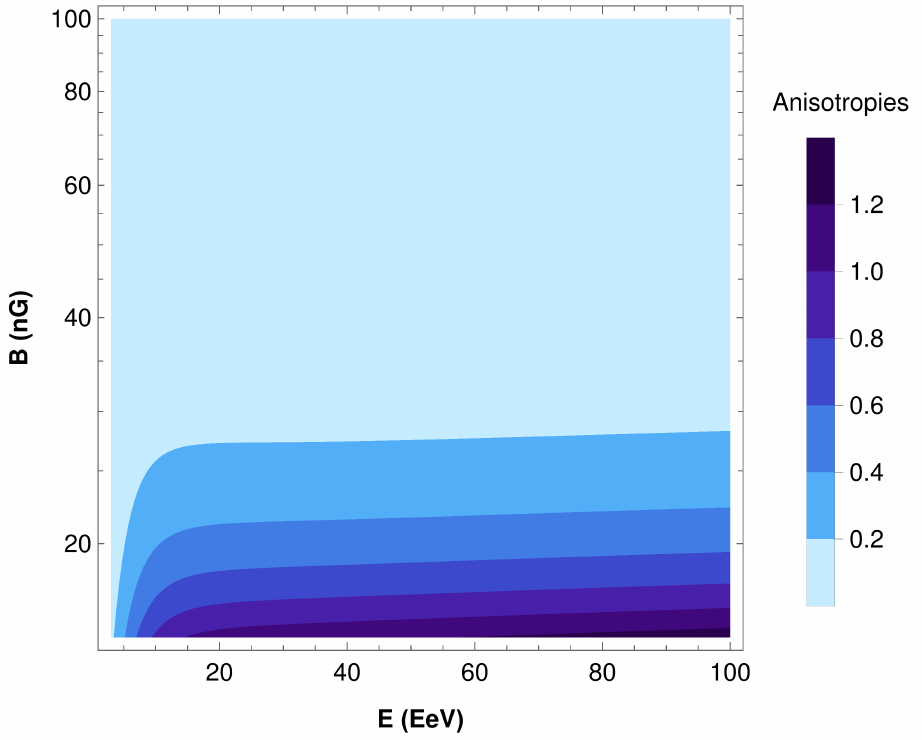} 
\includegraphics[scale=0.38]{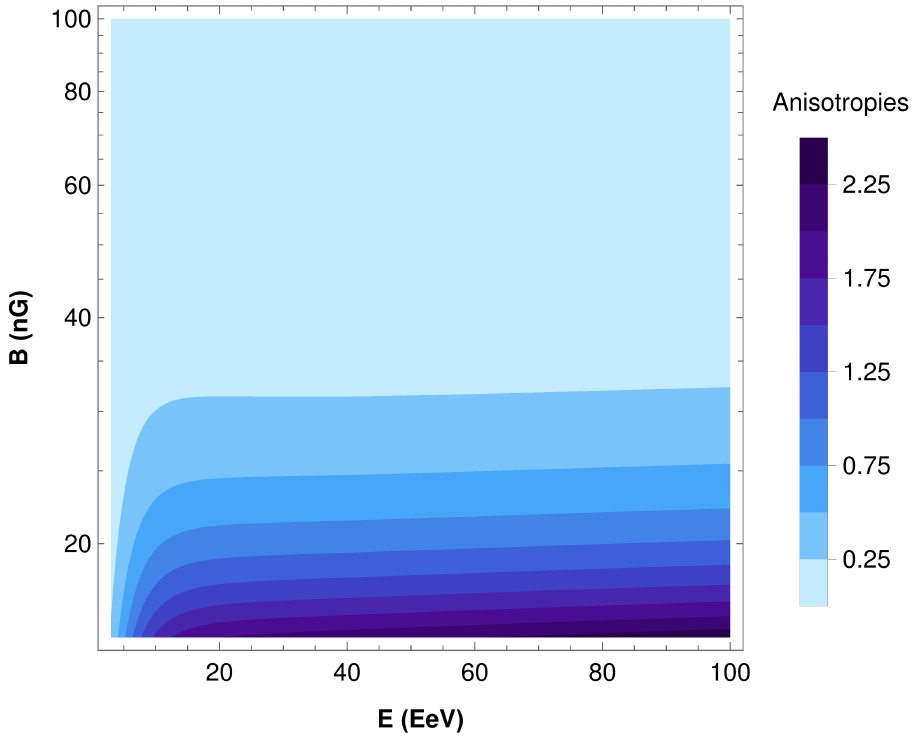}
\includegraphics[scale=0.38]{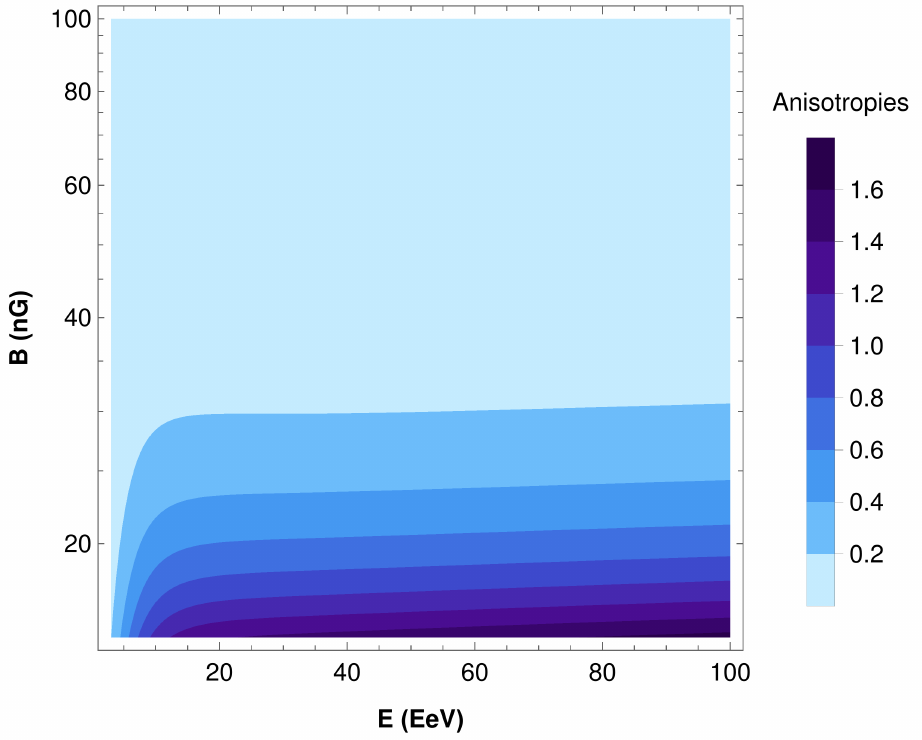}}
\caption{Variation of UHECRs' anisotropy as a function of the energy and the 
wide range of magnetic field strengths within a 1 Mpc thick plane for the $\Lambda$CDM model 
(left panel), $f(R)$ power-law model (middle panel) and the Starobinsky 
model (right panel).}
\label{fig5}
\end{figure}

%\section{Another modelling of magnetic field}\label{secIV}
Finally, for a general picture of the effect of magnetic field on
the anisotropy of UHECRs, we consider a scenario, where UHECRs propagate 
through a TMFs of $10-100$ nG within a $1$ Mpc thick plane. Fig.~\ref{fig5}
shows the results of this scenario. We observed that the level of anisotropy is high 
 for the low value of magnetic fields and low 
($< 0.2$) for the high value of 
magnetic fields. The low-level anisotropy is almost the same for all three 
cosmological models. The low-level anisotropies dominate most of the 
range we considered. The effect of the cosmological model for low-value anisotropy is less as compared to the cases 
we have discussed earlier. These results are due to the fact that the small 
source distance and the strength of the magnetic fields dominate the effects 
on the anisotropy of UHECRs.

\section{Summary and conclusion}\label{secIV}
\begin{figure}[h!]
\centerline{
\includegraphics[scale=0.6]{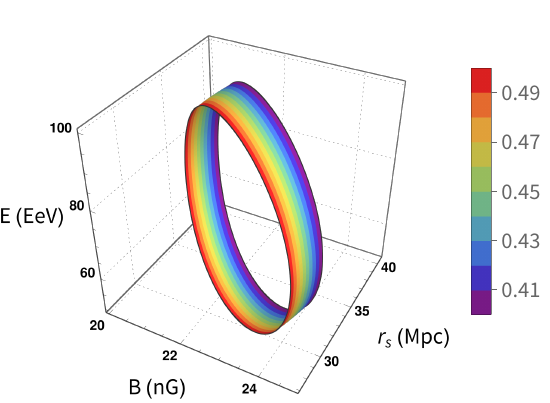} \hspace{1pt}
\includegraphics[scale=0.6]{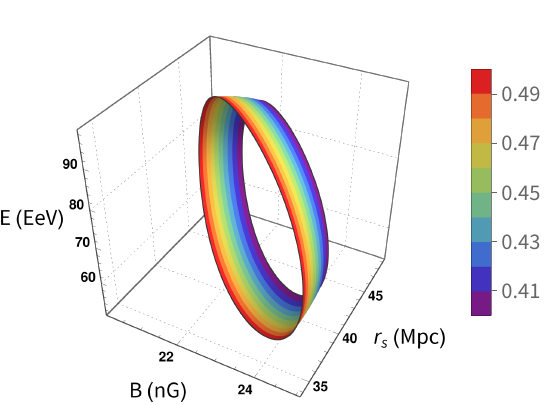} \hspace{1pt}
\includegraphics[scale=0.6]{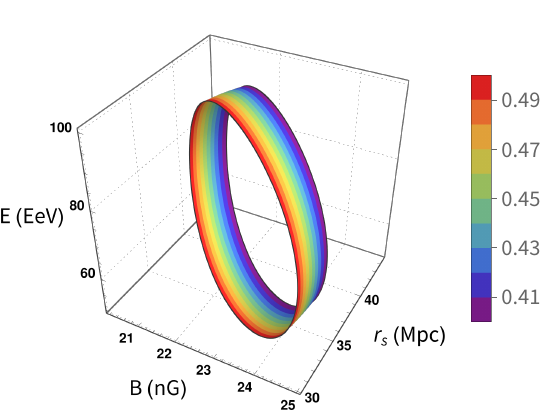}}
\caption{Variation of UHECRs' anisotropy with respect to energy, magnetic field
strength, and source distance according to the $\Lambda$CDM model, $f(R)$
power-law model and the Starobinsky model.}
\label{fig6}
\end{figure}

In this study, we considered in detail the diffusion of charged particles in 
TMFs and calculated the dipolar anisotropies of CRs in the energy range from 
0.1 EeV 100 EeV for the $\Lambda$CDM model, $f(R)$ power-law model and the 
Starobinsky model under the assumption of pure proton as well as pure helium, 
carbon and iron nuclei composition of UHECRs. 
We found that the anisotropy for the $\Lambda$CDM model and the Starobinsky 
model are almost identical to each other. On the other hand, the $f(R)$
power-law shows the highest anisotropy whether it is for protons and other 
nuclei. In the iron nuclei case, the lower range of dipolar amplitude 
dominated the plots in all those cosmological models considered here. To 
summarize the dependency of various factors such as energy, magnetic
field, source distance, and also the cosmological model on the UHECRs 
anisotropy, we draw the slice contour plots in Fig.\ \ref{fig6} by keeping a 
range of anisotropy for all these models from 0.4 to 0.5. From these plots, we 
can conclude that for the higher value of anisotropy, the required conditions 
are: higher energy, weak magnetic field strength, and closer to the source 
distance. These conditions are the results of the facts that high 
energetic CRs from a source at a particular distance are deflected less by a 
weak magnetic field than a strong magnetic field. Similarly, CRs from a nearby 
source with a particular energy have been deflected less by a magnetic field 
of a given strength than CRs from a distant source but with the same energy at 
the source. Thus, high energetic UHECRs from a nearby source in a weak magnetic 
field will be deflected very little from the source direction during their 
propagation, which leads to higher anisotropic distribution of UHECRs. For the same value of anisotropy with the given values of energy 
and magnetic field, the source distance depends on the cosmological model. 
For example, the anisotropy amplitude of 0.5 has been obtained (for certain 
values of magnetic field and energy) at source distances of $\sim$ 37 
Mpc and $\sim$ 33 Mpc for the power-law model and the Starobinsky model 
respectively. While the same magnitude of anisotropy can be obtained at a 
source distance of $\sim$ 30 Mpc in the $\Lambda$CDM model. Overall, the level 
of anisotropies of UHECRs predicted by our considered $f(R)$ gravity models 
which are in comparison to the prediction of the $\Lambda$CDM model are found 
to be consistent with observations at EeV energies \cite{harari,harari2021}.

At this point, it is to be noted that as our nearby Universe (within 100 Mpc) 
does not have a uniform distribution of sources, some directions have more sources than others. It is anticipated that, when this anisotropy is taken into 
account (for 2MRS galaxy catalog), the CRs distribution's dipole amplitude will rise by a factor of 1.5 
to 2 \cite{harari}. The CR distribution's asymmetry is indicated by this dipole
amplitude, and as it grows, the CR distribution will become more lopsided. The
direction of the dipole amplitude is also predicted to be the same as the
direction of motion of the Local Group \cite{harari, erdogdu}, which is the galaxy cluster
that we are a part of. In comparison to the anisotropies that the
Compton-Getting effect would produce, the anisotropies in the CRs
distribution that are brought about by the anisotropies in the galaxy
distribution are substantially bigger \cite{kachelriess}. Thus the anisotropy 
in the local distribution of sources is predicted to cause the anisotropy in 
CRs distribution.

It is important to note that additional deflections of CRs, primarily 
due to the galactic magnetic field, can alter both the amplitude and direction 
of the CR dipole. This can also lead to the creation of higher-order 
multipoles in the distribution of arrival directions \cite{harari}. For 
energies below a few EeV, these effects could potentially diminish the 
amplitudes calculated in this study to some degree.

As mentioned earlier, since the main aim of this work is to understand 
the possible effects of $f(R)$ gravity theory on the anisotropic distribution
of UHECRs during their diffusive propagation process in TMFs, our results are 
not quantified in the sense of experimental data. However, the results of this 
work provide important information about the diffusion of UHECRs in the TMFs. 
The dipolar anisotropy observed in UHECRs provides a valuable window into the 
underlying cosmological and astrophysical processes governing their 
propagation. As future experiments will provide more precise data on UHECRs, 
our work may help to establish a framework for discerning the underlying 
cosmological model that best explains the observed dipolar anisotropy. In 
conclusion, our study advances the comprehension of UHECRs' dipolar anisotropy 
within the diffusive regime by examining its manifestation in the context of 
the $\Lambda$CDM model, the $f(R)$ power-law model, and the Starobinsky model. 
The diverse predictions offered by these models underscore the importance of 
refining our understanding of UHECRs' sources, propagation, and the fundamental 
nature of gravity through continued experimental observations and theoretical 
investigations. However, we made several assumptions for simplicity. 
We considered a single source of UHECRs at varying distances, which may not 
capture the complexity of multiple sources. The galactic magnetic field, 
which can significantly affect UHECRs' propagation, was not included in our 
model. We also assumed that UHECRs' propagate through turbulent magnetic 
fields in extragalactic space, without considering structured fields. These 
assumptions, while necessary for this study, could be relaxed in future work 
to provide a more comprehensive model of UHECRs' propagation and anisotropy.

The complexity of distinguishing the effects of different cosmological 
models on the anisotropy of UHECRs arises from several factors. The unknown 
sources of UHECRs and their distribution can introduce anisotropies that 
might be misattributed to the cosmological models. The strength and structure 
of the magnetic fields, which are not well known, can deflect UHECRs and cause 
anisotropies, making it challenging to separate these effects from those of 
the cosmological model. The composition of UHECRs, another source of 
uncertainty, can affect the observed anisotropy as different particles 
interact differently with the intergalactic magnetic fields and the cosmic 
microwave background. Despite these uncertainties, statistical methods and 
careful assumptions may allow us to estimate the relative contributions of 
these factors to the observed anisotropy, enabling us to infer the properties 
of the cosmological models, the magnetic fields, and the UHECR sources and 
compositions. Further observational data and theoretical developments will 
be crucial in refining these estimates and improving our understanding of 
UHECRs. This work can be extended by taking into consideration other MTGs 
as well as the multiple sources along with the experimental data for 
the quantified and realistic analysis to understand the anisotropies of 
UHECRs.

\section*{Acknowledgements} UDG is thankful to the Inter-University Centre for 
Astronomy and Astrophysics (IUCAA), Pune, India for awarding the Visiting Associateship 
of the institute.

%%%%%%%%%%%%%%%%%%%%%%%%%%%%%%%%%%%%%%%%%%%%%%%%%%%%%%%%%%%%%%%%%%%%%%%%%%%%%%%%%%%
%%%%%%%%%%%%%%%%%%%%%%%%%%%%%%%%%%%%%%%%%%%%%%%%%%%%%%%%%%%%%%%%%%%%%%%%%%%%%%%%%%%
%%%%%%%%%%%%%%%%%%%%%%%%%%%%%%%%%%%%%%%%%%%%%%%%%%%%%%%%%%%%%%%%%%%%%%%%%%%%%%%%%%%

\end{document}